\RequirePackage{fix-cm}
\documentclass[natbib,smallextended]{svjour3}       
\smartqed  
\usepackage{graphicx}
\usepackage{amsmath}
\usepackage{amssymb}
\usepackage{xcolor}

 \journalname{SSRv}
\begin{document}

\title{Enrichment of the hot intracluster medium: observations}

\titlerunning{Enrichment of the hot intracluster medium: observations}        

\author{F. Mernier        \and
        V. Biffi \and
        H. Yamaguchi \and 
        P. Medvedev \and   
        A. Simionescu \and
        S. Ettori \and
        N. Werner \and
        J. S. Kaastra \and
        J. de Plaa \and
        L. Gu
}


\institute{F. Mernier \at
              MTA-E\"otv\"os University Lend\"ulet Hot Universe Research Group, P\'azm\'any P\'eter s\'et\'any 1/A, Budapest, 1117, Hungary \\
              Institute of Physics, E\"otv\"os University, P\'azm\'any P\'eter s\'et\'any 1/A, Budapest, 1117, Hungary \\
              SRON Netherlands Institute for Space Research, Sorbonnelaan 2, 3584 CA Utrecht, The Netherlands \\
              \email{mernier@caesar.elte.hu}
           \and
           V. Biffi \at Physics Department, Astronomy Unit, Trieste University, v. Tiepolo 11, 34143 Trieste, Italy \\
           INAF, Observatory of Trieste, v. Tiepolo 11, 34143 Trieste, Italy\\
           \and H. Yamaguchi \at
           Institute of Space and Astronautical Science (ISAS), JAXA, 3-1-1 Yoshinodai, Chuo-ku, Sagamihara, Kanagawa 252-5210, Japan \\
           \and
           P. Medvedev \at 
           Space Research Institute of the Russian Academy of Sciences (IKI), 84/32 Profsoyuznaya Str, Moscow, 117997, Russia  \\
           \and A. Simionescu \at
           SRON Netherlands Institute for Space Research, Sorbonnelaan 2, 3584 CA Utrecht, The Netherlands \\
           Institute of Space and Astronautical Science (ISAS), JAXA, 3-1-1 Yoshinodai, Chuo-ku, Sagamihara, Kanagawa 252-5210, Japan \\
           Kavli Institute for the Physics and Mathematics of the Universe, The University of Tokyo, Kashiwa, Chiba 277-8583, Japan \\
           \and S. Ettori \at
 INAF, Osservatorio di Astrofisica e Scienza dello Spazio, via Pietro Gobetti 93/3, 40129 Bologna, Italy \\
INFN, Sezione di Bologna, viale Berti Pichat 6/2, I-40127 Bologna, Italy \\
           \and N. Werner \at
           MTA-E\"otv\"os University Lend\"ulet Hot Universe Research Group, P\'azm\'any P\'eter s\'et\'any 1/A, Budapest, 1117, Hungary \\
           Department of Theoretical Physics and Astrophysics, Faculty of Science, Masaryk University, Kotl\'a\v{r}sk\'a 2, Brno, 611 37, Czech Republic \\
           School of Science, Hiroshima University, 1-3-1 Kagamiyama, Higashi-Hiroshima 739-8526, Japan \\
           \and J. S. Kaastra \at
           SRON Netherlands Institute for Space Research, Sorbonnelaan 2, 3584 CA Utrecht, The Netherlands \\
 Leiden Observatory, Leiden University, PO Box 9513, 
           2300 RA Leiden, The Netherlands\\        
           \and J. de Plaa \at
           SRON Netherlands Institute for Space Research, Sorbonnelaan 2, 3584 CA Utrecht, The Netherlands \\
           \and L. Gu \at
           RIKEN High Energy Astrophysics Laboratory, 2-1 Hirosawa, Wako, Saitama 351-0198, Japan \\
        SRON Netherlands Institute for Space Research, Sorbonnelaan 2, 3584 CA Utrecht, The Netherlands    
}

\date{Received: 6 July 2018 / Accepted: 3 November 2018}

\maketitle

\begin{abstract}
Four decades ago, the firm detection of an Fe-K emission feature in the X-ray spectrum of the Perseus cluster revealed the presence of iron in its hot intracluster medium (ICM). With more advanced missions successfully launched over the last 20 years, this discovery has been extended to many other metals and to the hot atmospheres of many other galaxy clusters, groups, and giant elliptical galaxies, as evidence that the elemental bricks of life -- synthesized by stars and supernovae -- are also found at the largest scales of the Universe. Because the ICM, emitting in X-rays, is in collisional ionisation equilibrium, its elemental abundances can in principle be accurately measured. These abundance measurements, in turn, are valuable to constrain the physics and environmental conditions of the Type Ia and core-collapse supernovae that exploded and enriched the ICM over the entire cluster volume. On the other hand, the spatial distribution of metals across the ICM constitutes a remarkable signature of the chemical history and evolution of clusters, groups, and ellipticals. Here, we summarise the most significant achievements in measuring elemental abundances in the ICM, from the very first attempts up to the era of \textit{XMM-Newton}, \textit{Chandra}, and \textit{Suzaku} and the unprecedented results obtained by \textit{Hitomi}. We also discuss the current systematic limitations of these measurements and how the future missions \textit{XRISM} and \textit{Athena} will further improve our current knowledge of the ICM enrichment.
\end{abstract}


\section{Introduction: the origin of metals}
\label{sec:intro}

Since the first complete theories of stellar nucleosynthesis in the 1950's \citep{cameron1957,burbidge1957}, it is well established that chemical elements heavier than helium are almost entirely the result of thermonuclear processes in stars at various stages of their lifetimes. This has a profound impact on our way to consider astronomy, as it means that all these building blocks of matter -- necessary to the formation of new stars, rocky planets, and even life -- have been once synthesised in the core of stars and/or during supernova (SN) explosions. Because all these heavy elements, or metals, have different atomic masses, hence different binding energies, they do not all originate from the same stellar sources. Instead, whereas carbon (C) and nitrogen (N) are mostly produced by low-mass stars in their asymptotic giant branch (AGB) phase, elements of intermediate atomic number ($8 \le Z \le 30$) are primarily produced by core-collapse supernovae (SNcc) and Type Ia supernovae (SNIa) \citep[for recent reviews on stellar nucleosynthesis, see][]{nomoto2013,thielemann2018}. A large fraction of heavier elements ($Z > 30$), on the other hand, is thought to be produced via either the rapid neutron-capture process ($r$-process) mostly in neutron star mergers \citep[for a review, see][]{thielemann2017}, or the slow neutron-capture process ($s$-process) mostly in AGB stars \citep[for a review, see][]{kappeler2011}.

\subsection{Core-collapse supernovae (SNcc)}
\label{sec:SNcc}

\begin{figure}
\begin{center}
  \includegraphics[width=0.8\textwidth]{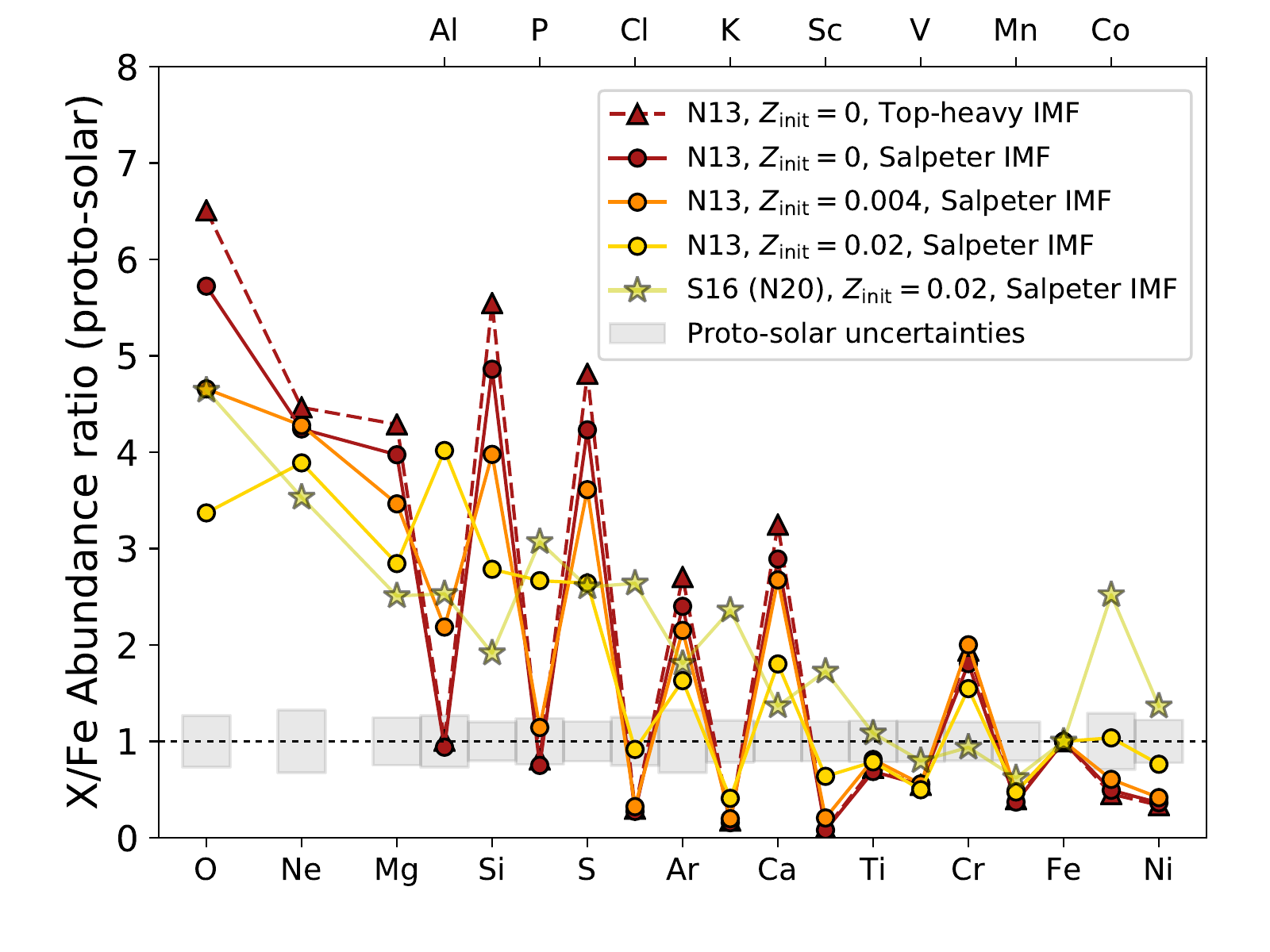} \\
  \includegraphics[width=0.8\textwidth]{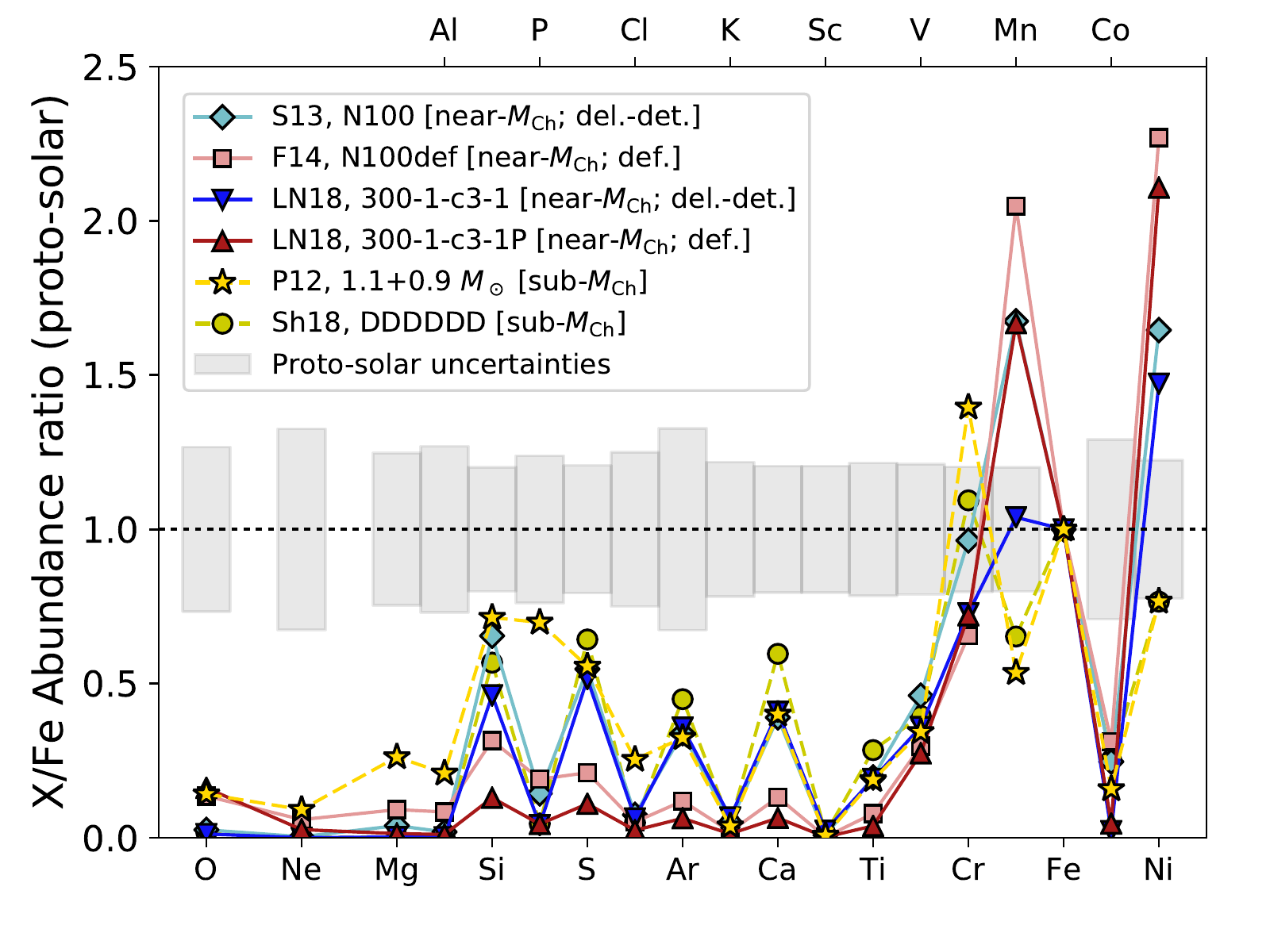} 
\end{center}
\caption{Abundance ratios predicted by SNcc and SNIa yield models from the literature. The abundances of F and Na are not shown here because they are mainly produced in AGB stars. \textit{Top}: SNcc yield models from \citet{nomoto2013} and \citet[][N20 model, including neutrino transport]{sukhbold2016}, assuming different values for $Z_\text{init}$. A comparison between different integrated IMFs -- "Salpeter" (slope index of -1.35) vs. "top-heavy" (slope index of -0.95) -- is also shown. \textit{Bottom}: SNIa yields. The N100 (delayed-detonation) and N100def (deflagration) models are calculated by \citet{seitenzahl2013} and \citet{fink2014}, respectively. The 300-1-c3-1 (delayed-detonation) and 300-1-c3-1P (deflagration) models are calculated by \citet{leung2018}. Whereas the four above models are near-$M_\text{Ch}$, the sub-$M_\text{Ch}$ models of \citet[][1.1+0.9$\,M_\odot$ WD merger]{pakmor2012} and \citet[][for 1 $M_\odot$, C/O=50/50, $Z_\text{init} = 0.02$, 12+16=1.0]{shen2018} are also shown.}
\label{fig:SN_models}       
\end{figure}

Massive stars ($\gtrsim 10\,M_\odot$) leave their main sequence when about 10\% of their hydrogen has burned into helium (He). In order to keep a balance between self-gravity and internal pressure, heavier elements are successively synthesised, then burned in turn, thereby forming concentric layers of burning elements in the stellar core. When the nuclear fusion of iron (Fe) and nickel (Ni) into heavier elements is reached, the process becomes endothermic. Consequently, their end-of-life nucleosynthesis stops at the Fe-peak elements, most of which remain locked in the collapsing core. In other words, SNcc eject mainly oxygen (O), neon (Ne), magnesium (Mg), silicon (Si), and sulfur (S), but very few amounts of heavier elements. The relative importance of these yields, however, is very sensitive to
\begin{enumerate}
\item the initial mass of the parent massive star;
\item the initial metallicity ($Z_\text{init}$) of the parent massive star;
\item details on how the explosion is driven.
\end{enumerate}
If one considers a simple stellar population instead of a single massive star, the yields should be integrated over the initial mass function (IMF) of such a population. In this case, assuming different IMFs -- e.g. either "Salpeter" \citep[i.e. a power-law with a slope index of -1.35;][]{salpeter1955} or "top-heavy" \citep[a shallower slope index of, e.g., -0.95;][]{arimoto1987} -- will produce different amounts of integrated yields.
Such a dependency of SNcc yield models on $Z_\text{init}$ and the IMF is illustrated in the upper panel of Fig.~\ref{fig:SN_models}, where various SNcc models adapted from \citet[][and references therein]{nomoto2013} and \citet{sukhbold2016} predict different X/Fe abundance ratios. Commonly used SNcc yield models from the literature include for example \citet{chieffi2004}, \citet{nomoto2006} and \citet{kobayashi2006} -- whose models were summarised and slightly updated in \citet[][]{nomoto2013} -- and \citet{sukhbold2016}. Recent updates have also been published by \citet{pignatari2016} and \citet{ritter2018}.

\subsection{Type Ia supernovae}
\label{sec:SNIa}

When a low-mass star ($\lesssim 10\,M_\odot$) ejects its upper layers at the end of its life, a dense core of degenerate matter remains (i.e. a white dwarf, WD). Unlike main-sequence stars, the internal pressure of WDs does not depend on their temperature. Consequently, if a WD gains sufficient mass, it may result into an explosive nucleosynthesis (starting from C) that will entirely disrupt it and eject even its heaviest produced elements. For this reason, in addition to Si and S, SNIa release large amounts of argon (Ar), calcium (Ca), chromium (Cr), manganese (Mn), Fe and Ni on one hand, and relatively small amounts O, Ne, and Mg on the other hand.

As an unsolved mystery, the precise nature of SNIa progenitors, as well as their subsequent explosion mechanism, is yet to be clarified \citep[for recent reviews, see e.g.][]{howell2011,maoz2012,maoz2014}. In fact, the explosive C-burning of the WD might result from either a gentle accumulation of material from a main-sequence companion (the single-degenerate scenario) or from a merger with another WD (the double-degenerate scenario). The former and latter scenarios are often associated with an explosive nucleosynthesis starting when the total WD mass is, respectively, close to (near-$M_\text{Ch}$) and well below (sub-$M_\text{Ch}$) the Chandrasekhar mass. This whole picture, however, may be more complicated as in some cases near-$M_\text{Ch}$ SNIa may also be considered in the double-degenerate scenario while sub-$M_\text{Ch}$ SNIa may also be single-degenerate \citep[e.g.][]{nomoto2013}.

In the near-$M_\text{Ch}$ case, the burning flame in the WD may either propagate always subsonically -- referred to as deflagration explosion -- or start subsonically before reaching supersonic velocities when propagating below a specific density -- referred to as delayed-detonation explosion \citep[for a recent review, see][]{nomoto2018}. In the sub-$M_\text{Ch}$ case, a violent WD-WD merger is thought to trigger an always supersonic burning flame (referred to as detonation explosion). Interestingly, all these different explosion mechanisms provide different yields, as illustrated in the lower panel of Fig.~\ref{fig:SN_models}. In fact, while deflagration models produce Ni in larger quantities, delayed-detonation models produce less Ni but more intermediate elements due to the incomplete burning stages in the SNIa explosion. Commonly used SNIa yield models from the literature include for example \citet[][near-$M_\text{Ch}$, 1D calculations]{iwamoto1999}, \citet[][near-$M_\text{Ch}$, 2D calculations]{maeda2010}, \citet[][sub-$M_\text{Ch}$, 1.1+0.9$\,M_\odot$ WD merger]{pakmor2012}, \citet{seitenzahl2013} and \citet[][near-$M_\text{Ch}$, 3D calculations]{fink2014}, \citet[][near-$M_\text{Ch}$, 2D calculations, with an extra dependence on the progenitor initial metallicity]{leung2018}, and \citet[][sub-$M_\text{Ch}$, detonation of one of the two WDs in the context of "dynamically-driven double-degenerate double-detonation" -- or DDDDDD -- scenario]{shen2018}.

\subsection{Asymptotic giant branch stars}
\label{sec:AGB}

Most metals lighter than O -- in particular C and N -- as well as fluorine (F) and sodium (Na) are predominantly synthesised by low-mass stars at the end of their life, i.e. during their asymptotic giant branch (AGB) phase \citep[e.g.][]{karakas2010}. Such elements are then easily released into the surrounding interstellar medium (and beyond) via powerful winds generated by these stars. As for SNcc, elemental yields originating from AGB stars are sensitive to the initial mass (or IMF) and metallicities of their stellar progenitors.

\section{Detecting metals (and measuring their abundances) in the ICM}
\label{sec:abundance_studies}

The presence of metals extends well beyond galactic scales, and even reaches the largest gravitationally bound structures of the Universe that are galaxy clusters. Specifically, the hot ($\sim$10$^6$--10$^8$ K), optically thin plasma (or intracluster medium, ICM\footnote{Formally, "ICM" should be restricted to the hot gas pervading galaxy clusters only. For convenience, however, in the following we use this acronym to designate also the hot atmospheres pervading galaxy groups and isolated, massive ellipticals.}) pervading galaxy clusters, groups, and giant elliptical galaxies is rich in chemical elements, which can be detected at X-ray energies via their emission lines \citep[for recent reviews, see][]{werner2008,boehringer2010,deplaa2013,deplaa2017a}.

\subsection{Spectral lines and abundance diagnostics}
\label{sec:lines_abundances}

Diagnostics of elemental abundances in the ICM are based on measurements of the flux in spectral lines as compared with that of the continuum.
Specifically, there are three processes that are responsible for the formation of thermal continuum radiation of any plasma: (i) bremsstrahlung emission (free-free), (ii) radiative recombination (free-bound), and  (iii) two-photon emission  \citep[see][for a review]{kaastra2008}.  At typical temperatures of a hot ICM plasma, the main contribution to the X-ray continuum spectra is made by the bremsstrahlung emission (free-free process).  Indeed, at temperatures above 1 keV,  radiation produced via electrons scattering off of protons and He nuclei  dominates the whole X-ray range \citep[see e.g.][]{mewe1986}.
Thus the emissivity (the emitted energy per unit time, energy and volume)  of the continuum radiation is given by
\begin{eqnarray}
\epsilon_\text{cont}(E) \sim \epsilon_{ff} \propto n_e \, T^{-1/2} \, e^{-E/kT} \sum\limits_{i} n_i\, Z_i^2 \, g_{ff} \nonumber \\
\sim n_e^2 \, T^{-1/2} \, e^{-E/kT} \, (1+4x)/(1+2x),
\label{eq:cont}
\end{eqnarray}
where $T$ and $n_e$ are the electron temperature and density, and $E$ is the energy of the emitted photon.
The factor $g_{ff} (T,Z_i,E)$ is a dimensionless quantity of order of unity (the Gaunt factor).
The sum is over all ions (with density $n_i$ and charge $Z_i$ of an ion)  present in the ICM. The last expression in Eq.~\ref{eq:cont} assumes that the sum is dominated by  hydrogen and helium  and  $n_e \approx n_p + 2 n_{{\rm He}} = n_p (1 + 2x)$, where $x = n_{He}/n_p$ stands for the helium-to-hydrogen ratio.

On the other hand, most X-ray lines are usually excited by collisional excitation by electrons. The integrated emissivity due to a collisionally excited line is given by

\begin{eqnarray}
\int \epsilon_\text{line}(E) dE \propto n(X^i)\, n_e\, E\, T^{-1/2}\, \Omega(T)\, e^{-\Delta E / kT} \nonumber \\
\propto \left[\frac{n(X^i)}{n_X}\right]\, \left[\frac{n_X}{n_p}\right]\, \left[\frac{n_p}{n_e}\right]\, n_e^2\, E\, \Omega(T)\, T^{-1/2}\, e^{-\Delta E/kT},
\label{eq:line}
\end{eqnarray}
where $\Delta E$ is the excitation energy above the ground state of the excited level and $\Omega(T)$ is the collision strength, which
usually varies only weakly with temperature \citep[see e.g.][for a review]{kaastra2008}.
Since the ICM is in, or very close to collisional ionisation equilibrium (CIE),  the ionization fractions -- $\left[n(X^i)/n_X\right]$ for an element $X$ -- depend only on the electron temperature $T$, and are independent of the ICM density.
Consequently,
the emissivity of a line is proportional to the square of the density and to the abundance of the relevant  element  $\left[n_X/n_p\right]$.  
At typical ICM (low) densities, the ratio of the line emission to the bremsstrahlung continuum is thus independent of the density, but keeps a direct proportionality to the abundance of the corresponding element. 
Since the electron temperature $T$ of the ICM can be derived from line ratios or the shape of X-ray continuum spectrum,
this line-to-continuum ratio (namely, the equivalent width of the line) can be easily converted into an elemental abundance.

In practice, elemental abundances in the ICM of clusters, groups, and ellipticals can be measured by fitting their X-ray spectra with models of CIE emitting plasma. The two spectral codes that are commonly found in the literature are AtomDB\footnote{http://www.atomdb.org} \citep[used in the model \textsc{APEC} as part of the XSPEC fitting package;][]{foster2012} and \textsc{SPEXACT}\footnote{https://www.sron.nl/astrophysics-spex} \citep[as part of the SPEX fitting package;][]{kaastra1996}. A detailed discussion on the uncertainties related to these two codes (and their subsequent atomic databases) is addressed in Sect.~\ref{sec:atomic_biases}.

Since $\left[n_X/n_p\right]$ is of the order $10^{-4}$--$10^{-7}$ for elements typically detected in the ICM, it is convenient to report abundances with respect to the Solar value. Here we choose to normalize all quoted abundances in the proto-solar units of \citet{lodders2009}, unless stated otherwise.

\subsection{Metals in the ICM: a brief history}
\label{sec:metals_history}

\begin{figure}
  \includegraphics[width=0.4\textwidth]{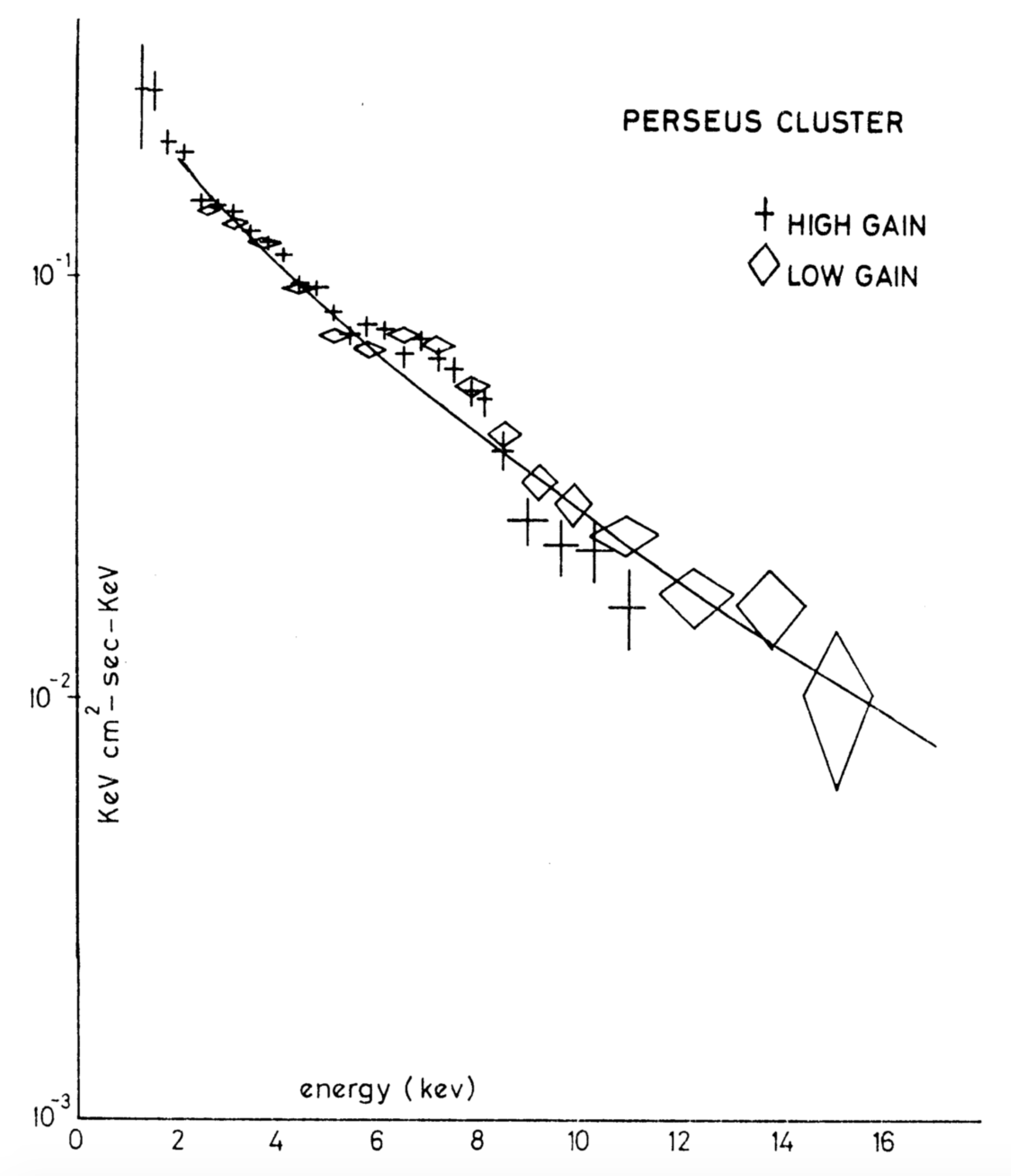} 
  \includegraphics[width=0.6\textwidth]{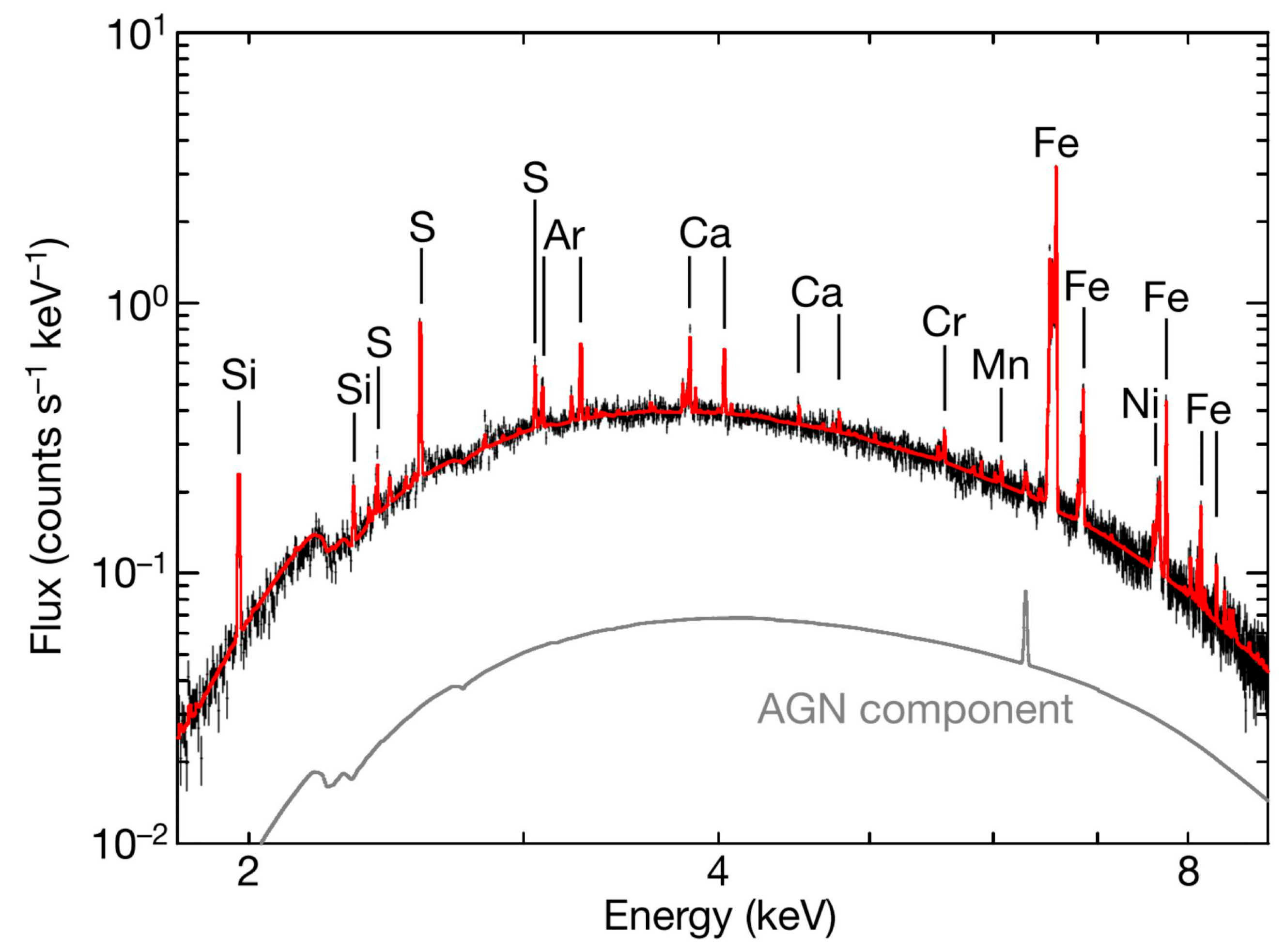} 
\caption{\textit{Left} \citep[from][reprinted with permission]{mitchell1976}: X-ray spectrum of the Perseus cluster core seen by the MSSL collimated proportional counter onboard Ariel V. \textit{Right} \citep[from][reprinted with permission]{THC2017}: X-ray spectrum of the Perseus cluster core seen by the SXS instrument onboard \textit{Hitomi}.}
\label{fig:Perseus_lines}       
\end{figure}

The first spectral signature of metals in the ICM was detected more than four decades ago, via the Fe-K emission line complex seen in nearby clusters by Ariel V \citep[][Fig.~\ref{fig:Perseus_lines}, \textit{left}]{mitchell1976} and OSO-8 \citep{serlemitsos1977}. Later on, the \textit{Einstein} observatory allowed the detection of other elements, such as O \citep{canizares1979}, Si \citep{mushotzky1981}, Mg, S, and Ar \citep{lea1982}. Despite these important detections, the breakthrough in abundance studies came with the \textit{ASCA} observatory. In addition to providing the first hints of the detection of Ca and Ni, \textit{ASCA} also allowed a first robust constraint of the Si, S, and Fe abundances \citep{mushotzky1996,fukazawa1998,matsushita2000,baumgartner2005}.

The next generation of X-ray observatories, including \textit{Chandra}, \textit{XMM-Newton} and \textit{Suzaku}, considerably increased the accuracy of these measurements \citep[see e.g.][]{boehringer2001,molendi2001a,finoguenov2002,buote2003,sanders2006b,werner2006a,deplaa2006,matsushita2007,deplaa2007,rasmussen2007,simionescu2009,degrandi2009,bulbul2012a,sasaki2014,konami2014,mernier2015,mernier2016a,thoelken2016}. 
The Reflection Grating Spectrometer (RGS) instrument onboard \textit{XMM-Newton} allowed to formally identify C \citep{werner2006b}, N, and Ne \citep{xu2002} emission lines in the ICM. In addition, the CCD-type European Photon Imaging Cameras (EPIC, onboard \textit{XMM-Newton}) and X-ray Imaging Spectrometer (XIS, onboard \textit{Suzaku}) allowed significant detections and measurements of Cr \citep[][with $>$2$\sigma$, $>$4$\sigma$, and $>$8$\sigma$, respectively]{werner2006a,tamura2009,mernier2016a} and Mn \citep[][with $>$1$\sigma$ and $>$7$\sigma$, respectively]{tamura2009,mernier2016a} abundances in clusters for the first time.

To date, the most spectacular abundance measurements have decidedly been achieved in the core of the Perseus cluster thanks to the exquisite spectral resolution of the micro-calorimeter Soft X-ray Spectrometer (SXS) onboard the \textit{Hitomi} mission \citep[][Fig.~\ref{fig:Perseus_lines}, \textit{right}]{THC2017}. In particular, the Cr, Mn, and Ni emission lines could be better identified and their abundances could be constrained with unprecedented accuracies (see also Sect.~\ref{sec:SN_models_Hitomi}).
This \textit{Hitomi} observation, however, was performed with the gate valve unopened (which consists of a 262-$\mu$m Be filter) because it was still in the initial operation phase. This substantially attenuated soft X-ray photons, limiting the SXS bandpass to above $\sim$1.8 keV (Fig.~\ref{fig:Perseus_lines}, \textit{right}). 
Detection with micro-calorimeters of the lighter elements, such as C, N, O, Ne, and Mg, as well as Fe-L emission is expected to be achieved with future observatories such as the X-ray Imaging and Spectroscopy Mission (\textit{XRISM}, formerly \textit{XARM}) and the Advanced Telescope for High-ENergy Astrophysics (\textit{Athena}) (see Sect.~\ref{sec:conclusion}).

Regrettably, the K-shell transitions of elements heavier than Ni (presumably produced by $r-$ and $s-$processes; see Sect. \ref{sec:intro}) occur at higher energies ($E > 10$ keV), which are not accessible with sufficient spectral resolution using current and upcoming X-ray telescopes. Therefore, this review does not cover these nucleosynthesis channels in further detail.


\section{Distribution of metals}
\label{sec:distribution}

Because the presence of chemical elements beyond galaxies is the signature of a direct interaction between sub-pc (stars and SNe) and Mpc (galaxy clusters) astrophysical scales, investigating the spatial distribution of metals in the ICM can reveal invaluable information on several aspects: (i) at which cosmic epoch, (ii) from which astrophysical locations, and (iii) via which transport process(es) clusters and groups were chemically enriched. Perhaps even more importantly, this allows to better understand the interplay between formation, growth, kinetic and thermal feedback, and metal enrichment of the largest gravitationally bound structures in the Universe. Because, presumably, enrichment via SNIa and SNcc are two distinct components occurring on different time scales after a starburst event, probing the distribution of both Fe and the other elements is of high interest.

Whereas observations alone can provide interesting constraints on the above questions, comparing them to cosmological hydrodynamical simulations is crucial to complete the picture. This aspect is reviewed in detail in the companion review by \citet[][in this topical collection]{biffi_review}.

\subsection{Central abundance distribution}
\label{sec:radial_core}

\begin{figure*}
  \includegraphics[trim=0.3cm 0cm 0.3cm 0cm, clip=true, width=1.0\textwidth]{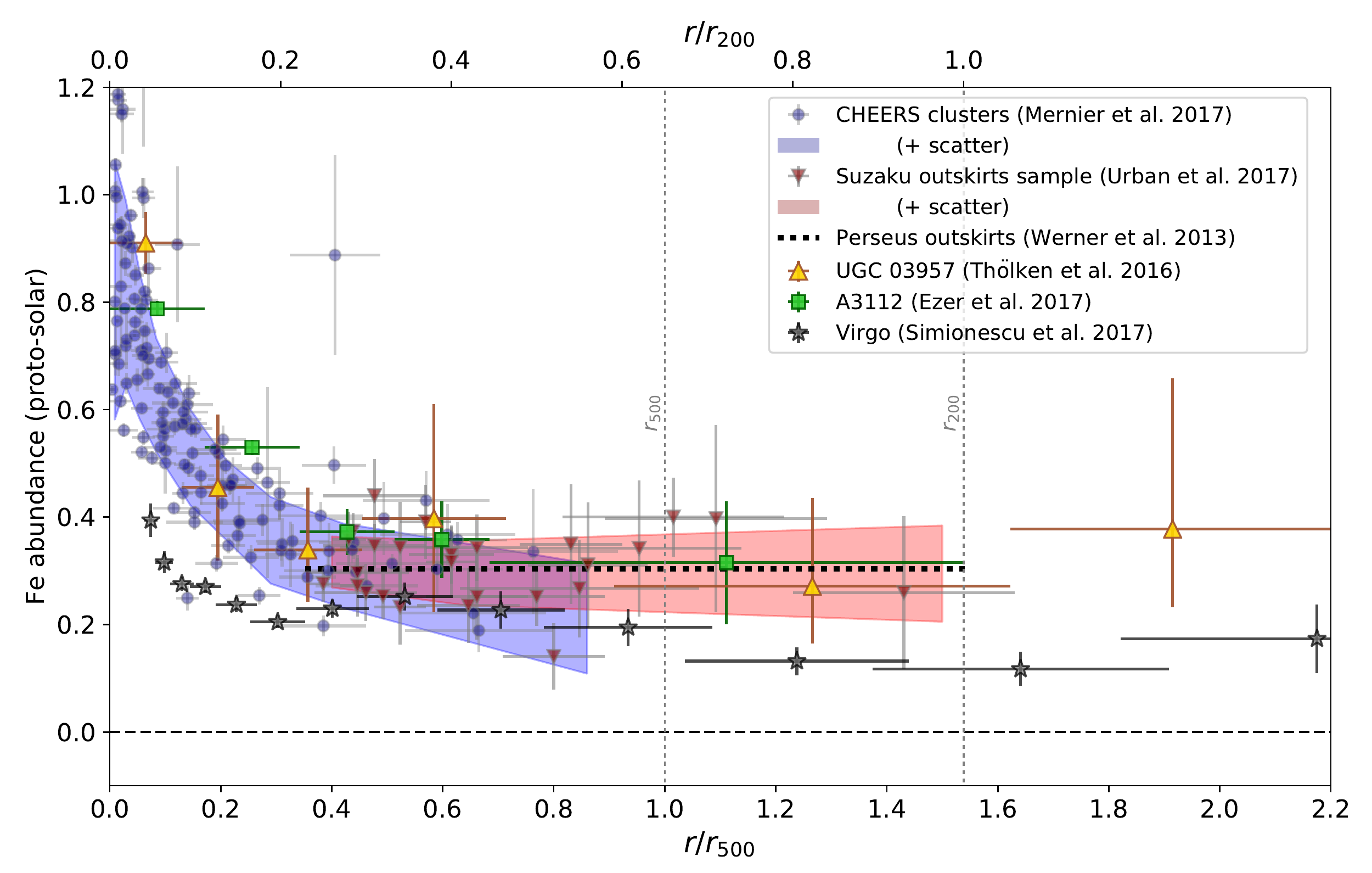}
\caption{Measured radial Fe abundance profile in cool-core clusters compiled from recent works. The conversion $r_{500} \simeq 0.65\,r_{200}$ is adopted from \citet{reiprich2013}. The blue circles and the blue shaded area show respectively the \textit{XMM-Newton} EPIC measurements of 23 clusters from the CHEERS sample \citep{mernier2017} and the \textit{Suzaku} XIS measurements in the outskirts of 10 clusters \citep{urban2017}. From the same respective studies, the red circles and the red shaded area show the intrinsic scatter of the measurements \citep[following the method of][]{mernier2017}. The black dotted line, the yellow triangles, the green squares, and the black stars show respectively the \textit{Suzaku} XIS measurements of the outskirts of the Perseus cluster \citep[][averaged value]{werner2013}, the UGC\,03957 group \citep{thoelken2016}, the A\,3112 cluster \citep{ezer2017}, and the Virgo cluster \citep[][azimuthally averaged along the N, S, and W arms]{simionescu2017}. All the abundances are rescaled with respect to the proto-solar values of \citet{lodders2009}.}
\label{fig:radial_Fe}       
\end{figure*}

The \textit{ASCA} observatory was the first satellite that made possible the spatial investigation of metallicity in the bright cool-cores of relaxed nearby clusters. Early observations of the Centaurus cluster \citep{allen1994,fukazawa1994} revealed a decreasing radial gradient of Fe from the centre to the outskirts. Later on, observations of larger samples using \textit{BeppoSAX} \citep[][]{degrandi2001} and \textit{XMM-Newton} \citep{degrandi2004} established that central Fe peaks are found in cool-core clusters\footnote{Cool-core clusters are usually characterised by a centrally peaked surface brightness, together with a stratified ICM entropy and a cooling time that is shorter than the Hubble time \citep{molendi2001b}.} while non-cool-core clusters exhibit significantly flatter central profiles \citep[][see also the recent study by]{lovisari2019}. Since these first results, such gradients have been routinely measured in cool-core clusters, groups, and ellipticals \citep[e.g.][]{buote2003,deplaa2006,rasmussen2007,leccardi2008,sanderson2009,million2010,lovisari2011,matsushita2011,sasaki2014,mernier2015,sanders2016,mernier2017,lovisari2019}, and most of the efforts have been devoted to the interpretation of central Fe peaks in cool-core systems. A compilation of recent radial measurements of Fe in cool-core clusters -- including the central peak extending out to $\sim$0.5$\,r_{500}$ can be seen in Fig.~\ref{fig:radial_Fe}. Measurements within this radius are taken from \citet{thoelken2016}, \citet{ezer2017}, \citet{mernier2017}, and \citet{simionescu2017}.

At first glance, such central Fe peaks in cool-core clusters were naturally thought to be explained by the low-mass stellar population from the central brightest cluster galaxy (BCG) as, unlike our Milky Way, the latter is usually red-and-dead with very few massive stars at present times. In fact, the metal mass estimated to be ejected from BCGs was found to correlate with the amount of Fe of the central ICM phase \citep{degrandi2004}. In parallel, pioneer observations of the radial distribution of elements rather produced by SNcc \citep[O, Mg, Si;][]{boehringer2001,tamura2001,finoguenov2002,matsushita2003, werner2006b} reported flatter profiles than for Fe. This result naturally suggested that the Fe peak would originate from delayed (and still ongoing) SNIa in the central cD galaxy, while the bulk of the SNcc enrichment occurred earlier and mixed more efficiently in the ICM \citep{boerhinger2004}. A radial increase of the Si/Fe ratio was also reported by \textit{Chandra} in galaxy groups \citep{rasmussen2007,rasmussen2009}, strengthening that initial idea of a secular evolution of the SNIa enrichment with respect to the SNcc enrichment.

With time, however, the improvement of instrumental calibrations combined with a better understanding of the background gradually revealed a completely different picture of the central ICM enrichment. Several recent studies on individual systems indeed reported a central peak not only in Fe-peak elements, but also in elements produced predominantly by SNcc \citep{deplaa2006,sato2009a,sato2009b,simionescu2009,murakami2011,bulbul2012a,mernier2015}, thereby challenging the interpretation of a late central enrichment coming from the SNIa explosions of the BCG. This was further confirmed by \citet{mernier2017}, who investigated the abundance radial profiles of 44 cool-core systems -- the CHEmical Enrichment Rgs Sample (CHEERS), including clusters, groups, and massive ellipiticals -- and showed that their azimuthally averaged SNIa-to-SNcc contribution remains remarkably uniform out to at least $\sim$0.5$\,r_{500}$. The facts that, even in the very center of cool-core systems, the abundance ratios remain very similar to that of the Milky Way \citep[Sect. \ref{sec:SN_models}; see also][]{deplaa2017b,THC2017,simionescu2018,mernier2018b} and that the abundance pattern measured by \textit{Hitomi} within and offset of the core were found to be similar within uncertainties \citep{simionescu2018} further support this picture.

This surprising similarity between the distribution of SNIa and SNcc products for apparently very different systems looks more elegant to report but is in fact less trivial to interpret. Two scenarios may compete:
\begin{itemize}
\item either the late, ongoing central enrichment from the BCG (assumed to be associated with the central abundance peaks) is produced by \emph{both} SNIa and SNcc -- or alternatively, by extremely $\alpha$/Fe-rich stellar winds;
\item or the early-type stellar population (and their consequent SNIa explosions) from the BCG does not contribute significantly to the central enrichment of cool-core systems. In this case, the origin of the central abundance peaks may not be related to the current stellar content of the BCG.
\end{itemize}

The recent mounting evidence that the bulk of the central Fe peak was already in place at $z \simeq 1$, before its radial extent broadens with cosmic time \citep[][see also Sect.~\ref{sec:redshift}]{degrandi2014,mantz2017} would support the BCG origin for the central enrichment. On the contrary, the moderate levels of recent star formation within local BCGs \citep[appearing red and dead at present times; e.g.][]{mcdonald2018} tends to disfavour the former scenario. A good compromise to this paradox would be the scenario of an early central enrichment (at $z>1$, i.e. before the Universe was half its age), occurring either \textit{in situ} (during the early BCG assembling via both important episodes of star formation and short delay time SNIa) or via the infall of already enriched, low entropy subhaloes towards the cluster core. Future detailed multi-wavelength observations coupled to chemodynamical simulations tracing the formation of BCGs would help to better understand this central enrichment picture in cool-core systems.

\subsubsection*{Central abundance drops}
\label{sec:drops}

While abundances are, on rather large scales, centrally peaked, several studies reported an inversion in the very core (i.e. within a few tens of kpc or less) of some systems. These central abundance drops were detected mostly in galaxy groups or giant ellipticals \citep[e.g.][]{rasmussen2007,rafferty2013,panagoulia2015,mernier2017,gendron-marsolais2017}, but sometimes also in more massive clusters \citep[e.g.][]{sanders2002,johnstone2002,churazov2003,churazov2004,million2010,mernier2017}. In some cases, these drops are simply an artifact resulting from a (too simplistic) single-temperature modelling for a multiphase plasma \citep[][see also Sect.~\ref{sec:other_biases}]{werner2006a}. In some other cases, however, these drops persist even when accounting for a more complex temperature structure \citep{panagoulia2013,panagoulia2015,mernier2017}, thereby requiring additional explanations.

Other fitting biases have been considered to explain this apparent lack of central abundances. For instance, as shown by the \textit{Hitomi} observations of Perseus \citep{THC2018_rs}, the effect of resonant scattering is the strongest at maximum surface brightness in the very core while usually ignored in spectral models \citep[see also][in this topical collection]{gu_review}. Nevertheless, \citet{sanders2006a} showed that such an effect could not entirely explain the abundance drops \citep[see also][]{gendron-marsolais2017}. The accurate spatial resolution of \textit{Chandra} also allows to discard possible contamination of the spectra by the central AGN \citep[e.g.][]{sanders2016} or by projection effects \citep{sanders2007}, hence suggesting that such abundance drops may be real. However, other spectral effects need to be further investigated \citep[e.g. incorrect assumptions on the central helium abundances and sedimentation,][see also Sect.~\ref{sec:sedimentation_biases}]{mernier2017}.

Another possibility would be that in the most central, coolest, and lowest entropy regions of the ICM, a significant fraction of metals deplete into dust (thus becoming not visible in the X-ray window), before getting uplifted and re-heated to the X-ray emission regime at larger radii via feedback processes from the central AGN \citep{panagoulia2015}. One key consequence of this scenario is that Ne and Ar should not exhibit central drops as noble gases are inefficient in getting depleted into dust. This prediction, however, might be in tension with measurements of the Ar profile in cool-core clusters \citep{mernier2017}.

\subsection{Metals in cluster outskirts}
\label{sec:radial_outskirts}

Cluster outskirts are undoubtedly a region of great interest as they contain most of the cluster volume and provide direct information on how the ICM forms, accretes and contributes to the growth of large scale structures \citep[][in this topical collection]{walker_review}.
Due to the low X-ray surface brightness of these outermost regions, however, metallicity measurements beyond one-half of their virial radii remain sparse. Arguably, the best abundance measurements at large radii so far have been provided by the \textit{Suzaku} satellite, which is less affected by particle background than \textit{XMM-Newton} and \textit{Chandra}. \citet{fujita2008} observed that the ICM between the merging clusters Abell\,399/401, close to their virial radii, is enriched to $\sim$0.3 of the proto-solar level. \textit{Suzaku} XIS observations of the Perseus cluster in 78 independent spatial bins and along 8 azimuthal directions revealed a uniform iron abundance of $Z_{\rm Fe} = 0.304\pm0.012$ proto-solar, as a function of both azimuth and radius, out to $r_{200}$ \citep{werner2013}. Very deep \textit{Suzaku} observations of Abell\,3112 \citep{ezer2017} as well as the analysis of archival \textit{Suzaku} data for the outskirts of ten massive clusters \citep{urban2017} confirm the results of the Perseus observations. \citet{urban2017} find that, across their sample, the Fe abundances are consistent with a constant value, $Z_{\rm Fe} = 0.306\pm0.012$ proto-solar, which is remarkably similar to the value measured for the Perseus cluster. Observations of the outskirts of the lower mass group UGC\,03957 with Suzaku also reveal a metallicity that is consistent with 0.3 Solar \citep{thoelken2016}. These measurements are summarised in Fig.~\ref{fig:radial_Fe}, where the gradual flattening of the Fe distribution beyond $\sim$0.5$\,r_{500}$ towards the uniform value of $\sim$0.3 Solar can be seen. The situation is slightly different for the Virgo cluster \citep[][see also Fig.~\ref{fig:radial_Fe}]{simionescu2017}, where the Fe profile flattens below $\sim$0.5$\,r_{500}$ and converges at a lower level, towards 0.2 Solar. Deeper observations with future instruments will help to clarify whether (and to which extent) the enrichment in Virgo is really different from other clusters and groups.

If the bulk of Fe was added at relatively recent times -- via e.g. ram-pressure stripping, the relative inefficiency of mixing (due to the early stratification of the entropy profile) would result in substantial radial gradients and azimuthal inhomogeneities of the metallicity distribution even in cluster outskirts \citep[e.g.][]{kapferer2006}. Instead, the observed homogeneous distribution of metals in the outskirts of clusters was among the first strong indications that most of the enrichment of the ICM took place \textit{before} the entropy gradient was established, hence \textit{before} cluster formation (at $z>2$--3, i.e. as early as, or even earlier than the central metal enrichment -- see Sect.~\ref{sec:radial_core}). Despite the potential problems and systematics pointed out by \citet{molendi2016}, the remarkable agreement between the Fe abundance measured in the outskirts of various clusters suggests that these measurements are reliable. Moreover, these observational results are in line with cosmological hydrodynamical simulations, which indicate a pre-enrichment scenario as well. In fact, recent simulations \citep{rasia2015,biffi2017,biffi2018}, predicting a uniform metallicity of $\sim$0.3 proto-solar, i.e. in remarkable agreement with the above observational results, show that pre-enriched gas is expelled from the shallow potential wells of early galaxies via AGN feedback and is distributed over large scales in the proto-cluster region, before being later accreted into the forming cluster. Further details on these predictions and their comparison with observations are discussed in \citet[][in this topical collection]{biffi_review}.

In addition, deep \textit{Suzaku} XIS observations of the Virgo cluster allowed abundance measurements for elements other than Fe at large radii, revealing also a uniform chemical composition throughout the cluster volume \citep{simionescu2015}. Specifically, these authors showed that the Mg/Fe, Si/Fe, and S/Fe ratios -- all reliably tracing the SNcc-to-SNIa contribution of the enrichment -- are remarkably flat out to $\sim$1.3$\,r_{200}$, confirming the previous hints reported by \citet{sasaki2014} for a sample of four galaxy groups (also observed with \textit{Suzaku} XIS). Furthermore, a pure SNcc enrichment (i.e. without SNIa contribution) could be firmly ruled out in the Virgo outskirts with high significance, even beyond half of the virial radius \citep{simionescu2015}. Combined with the results of, e.g., \citet{ezer2017} and \citet[][see also Sect.~\ref{sec:radial_core}]{mernier2017}, the emergent picture is that the ICM is enriched with a remarkably similar relative contribution of SNIa and SNcc products, from its very core out to the limits of its virialised regions.

\subsection{Inhomogeneities and redistribution of metals}
\label{sec:inhomogeneities}

Despite the remarkable average uniformity of metals reported in cluster outskirts, metals still act as passive tracers of gas motions, hence they are not expected to be distributed homogeneously everywhere in the ICM. In fact, measuring accurately their 2D spatial distribution in clusters and groups is valuable to better understand (i) their transport and diffusion processes from the interstellar medium (ISM) to the ICM, and (ii) the ICM (thermo-)dynamics in general. 

\begin{figure}
  \includegraphics[width=0.435\textwidth]{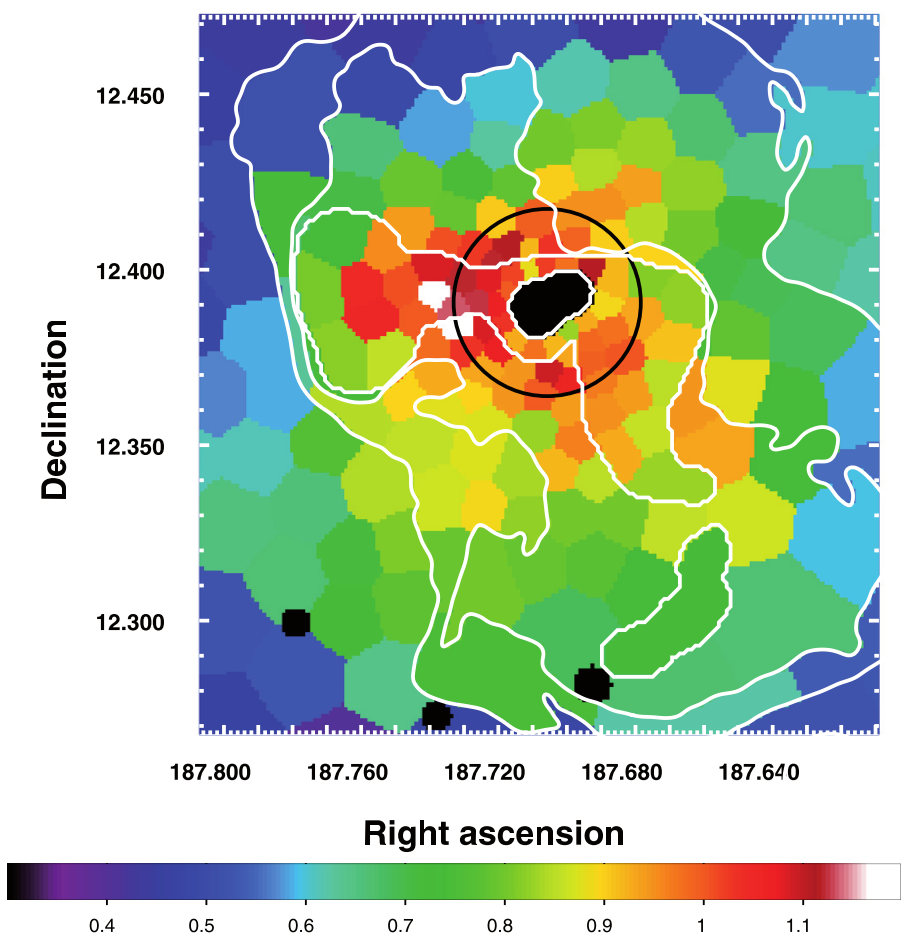} 
  \includegraphics[width=0.556\textwidth]{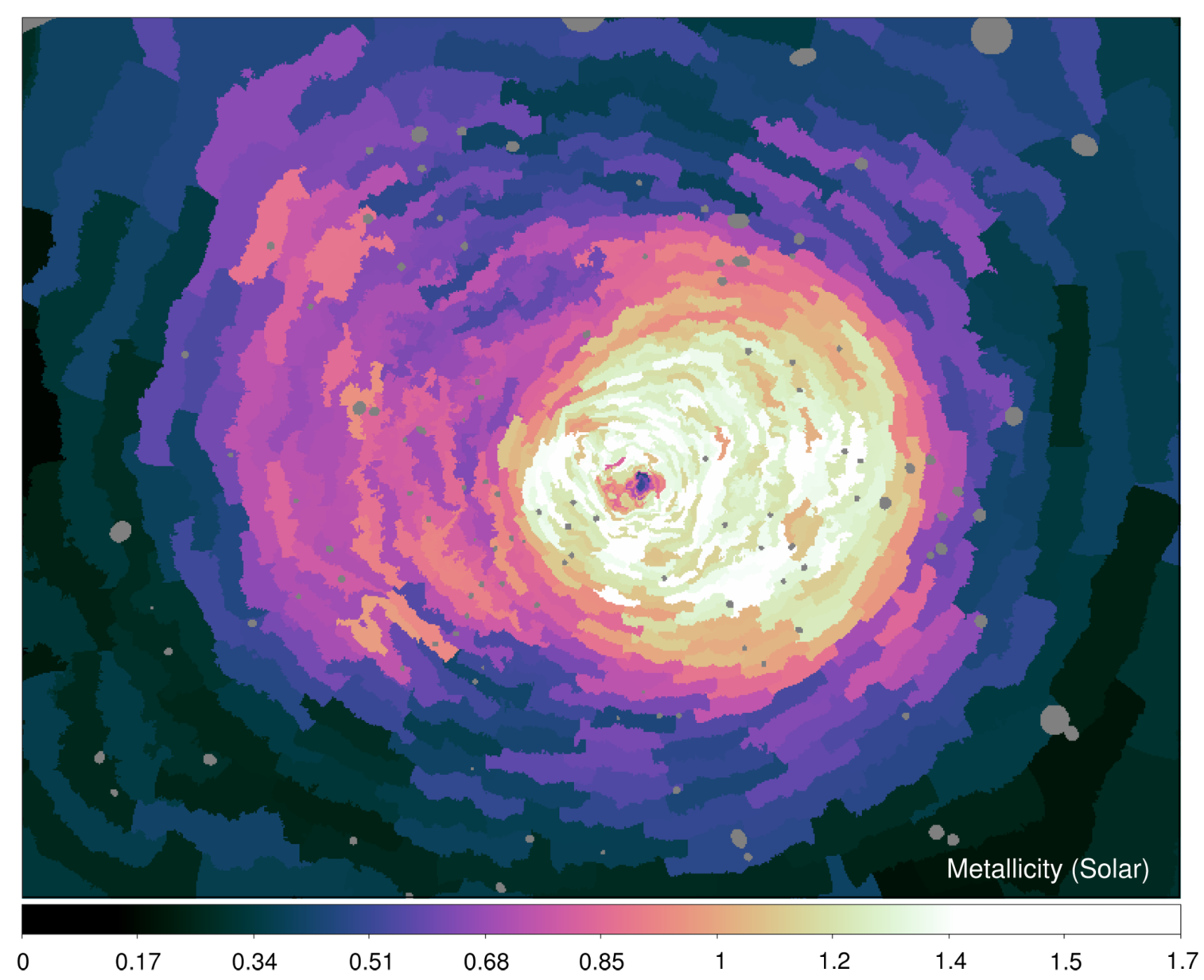} 
\caption{\textit{Left} \citep[from][reprinted with permission]{simionescu2008}: Fe metal map of the giant elliptical M\,87. Regions of most metal-rich gas coincide with the lobes of radio emission (white contours), suggesting an efficient uplifting by the AGN feedback. \textit{Right} \citep[from][reprinted with permission]{sanders2016}: Fe metal map of the Centaurus cluster, exhibiting a central abundance redistribution at larger scales via sloshing. A central drop in Fe can also be seen in the very core (see Sect.~\ref{sec:drops}).}
\label{fig:metal_maps}       
\end{figure}

Several processes may affect the distribution of metals in the ICM \citep[for a review, see][]{schindler2008}. Among the most spectacular examples, it was found that Fe follows remarkably well the (radio-emitting) jets of relativistic plasma ejected by the central AGN during its kinetic feedback mode. These jets, which are thought to shape X-ray cavities in cool-core systems and to provide a substantial source of re-heating to the radiatively cooling gas \citep[for a detailed review, see][in this topical collection]{werner2018}, are efficient at redistributing metals on both kpc \citep[e.g. M\,87;][see Fig.~\ref{fig:metal_maps} \textit{left}]{simionescu2008} and Mpc scales \citep{kirkpatrick2009,kirkpatrick2011,kirkpatrick2015}.

Another way to release metals from the ISM to the ICM could be achieved via ram-pressure stripping of gas-rich galaxies falling into dense cluster cores \citep[for good examples of ram-pressure stripped X-ray subhalo in a cluster, see e.g.][see also Simionescu et al. \citeyear{simionescu2019}, in this topical collection]{neumann2001,sun2010,eckert2014,ichinohe2015,degrandi2016,su2017}. This could be the case of A\,4059, in which an X-ray bright, cool, metal-rich blob was found significantly offset from the BCG and with no optical counterpart, thus perhaps originating from a neighboring late-type galaxy that may have lost its gas during its passage close to the BCG \citep{reynolds2008,mernier2015}.

At larger cluster-centric distances, gas sloshing may help to redistribute metals that were already present in the ICM. Across cold fronts generated by sloshing motions, which result from the encounter of a minor offset merger \citep[for a review, see][]{markevitch2007b}, the metallicity is observed to drop abruptly, in a comparable way to surface brightness and temperature discontinuities \citep[e.g.][Fig.~\ref{fig:metal_maps} \textit{right}]{simionescu2010,osullivan2014,ghizzardi2014,sanders2016}. This trend of the metallicity to follow the sloshing pattern of the gas suggests that these motions may redistribute metals within cold fronts, but are not efficient in mixing them with the ambient ICM at larger radii.

Finally, inhomogeneities in recent cluster mergers may point towards an effective mixing of metals during merging events \citep[e.g.][]{lovisari2011}. Detailed studies on this aspect are still challenging because the high temperature of such systems ($\sim$7--15 keV) lowers the metal line emissivities in the X-ray band.


\section{Chemical composition of the ICM and stellar nucleosynthesis yields}
\label{sec:SN_models}

As already mentioned in Sect.~\ref{sec:intro}, the abundance ratios of the yields produced by SNcc and SNIa provide invaluable information on their environmental conditions, their explosion mechanisms and/or their progenitors (Fig.~\ref{fig:SN_models}). Unfortunately, only a few tens of SN remnants are known in our Galaxy and the complicated physics of such plasmas make their abundances difficult to derive accurately \citep{vink2012,yamaguchi2014}. On the other hand, as seen in Sect.~\ref{sec:abundance_studies}, the ICM has been enriched by billions of SNIa and SNcc (thus providing a better representation of all the SNe in the Universe) and its abundances are much easier to constrain because the hot atmospheres pervading clusters, groups, and ellipticals are in CIE. Naturally, the past discovery of metals in the ICM opened an excellent opportunity to constrain SNcc and SNIa properties (and their relative contributions to the overall enrichment) by deriving the abundance ratios of different elements in this gas. In order to reach good statistics, and because metal emission lines are particularly prominent at moderate ICM temperatures, such studies are often limited to the central regions of cool-core clusters. 

As stated in Sect. \ref{sec:radial_core}, cluster galaxies (and particularly BCGs) are usually early-type and read-and-dead. Comparing the chemical composition (and the corresponding SNIa-to-SNcc enriching contributions) of large structures like galaxy clusters with that of our own Solar System is valuable also to understand our particular relationship to the chemical history of the Universe.

\subsection{Early results from previous (and ongoing) missions}
\label{sec:SN_models_previous}

A few years after the launch of \textit{ASCA}, the pioneer attempts to compare ICM abundances to nucleosynthesis models suggested an enrichment entirely coming from SNcc \citep{loewenstein1994,mushotzky1996,loewenstein1996}. Later studies, however, suggested that SNIa also contribute significantly to the enrichment \citep{fukazawa1998,finoguenov1999,finoguenov2000a,finoguenov2000b}. Overall, the global trend inferred from the \textit{ASCA} results was an increase of the SNcc contribution with the mass of the system \citep{fukazawa1998,baumgartner2005}. 

With the \textit{XMM-Newton} mission, the general picture somewhat changed and became better clarified. In particular, the abundance ratios in the ICM of 2A 0335+096 \citep{werner2006a} and S{\'e}rsic 159-03 \citep{deplaa2006} suggested a $\sim$25--50\% contribution of SNIa to the total enrichment, with no need for invoking an additional contribution from Population III stars. Moreover, the Ca/Fe ratio was found to be underestimated by the assumed SN yields. These results were later confirmed on a larger number of observations \citep[][see also Sato et al. \citeyear{sato2007} for results using \textit{Suzaku}]{deplaa2007}. Unlike previous \textit{ASCA} results, \citet{deplaa2007} and \citet{degrandi2009} found no dependency between temperature and the abundance ratios, suggesting that the SNIa and SNcc enrichment mechanism at play in massive and less massive clusters is very similar.
\citet{mernier2016a} used the CHEERS catalogue (44 cool-core systems using \textit{XMM-Newton} EPIC) to extend this conclusion to lower mass systems. Taking several sources of systematic uncertainties into account, they compiled the average X/Fe abundance pattern, globally representative of the ICM in nearby cool-core systems. In a second paper, and following the method of \citet{deplaa2007}, \citet{mernier2016b} compared these average ICM abundance ratios to various SNcc and SNIa yield models available in the literature \citep[e.g.][]{iwamoto1999,nomoto2013,seitenzahl2013,fink2014}. In addition to Ca/Fe (see above), the SNIa models also failed to reproduce the Ni/Fe ratio (found to be super-solar, though suffering from large uncertainties due to discrepancies between the MOS and pn instruments) if only one type of explosion is assumed. Above all, despite being the most comprehensive study to date on this aspect, it was demonstrated that systematic uncertainties on the abundance ratios clearly dominate over the statistical ones, stressing the need for higher energy resolution observations.

\begin{figure}
  \includegraphics[width=0.525\textwidth]{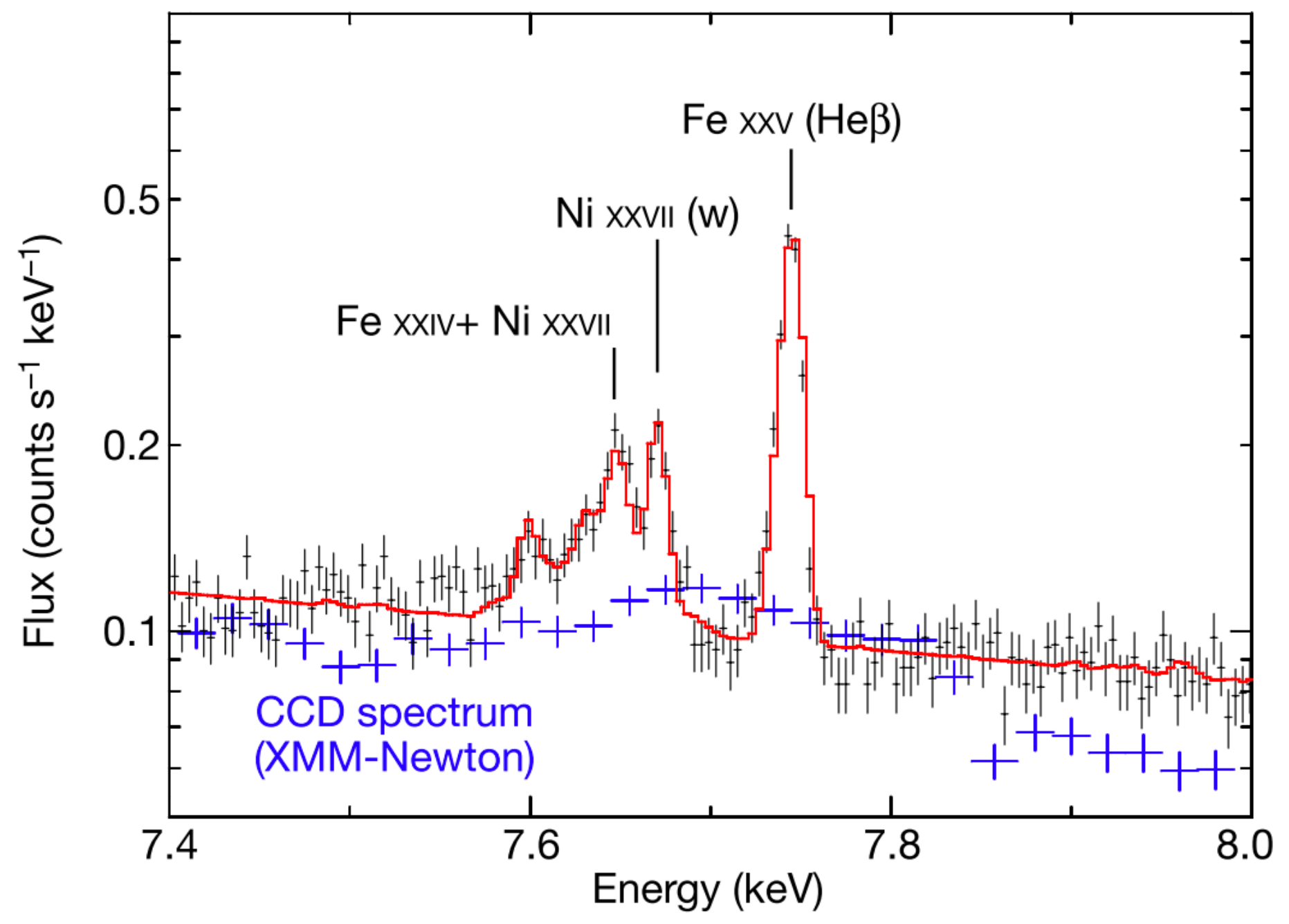}
  \includegraphics[width=0.47\textwidth]{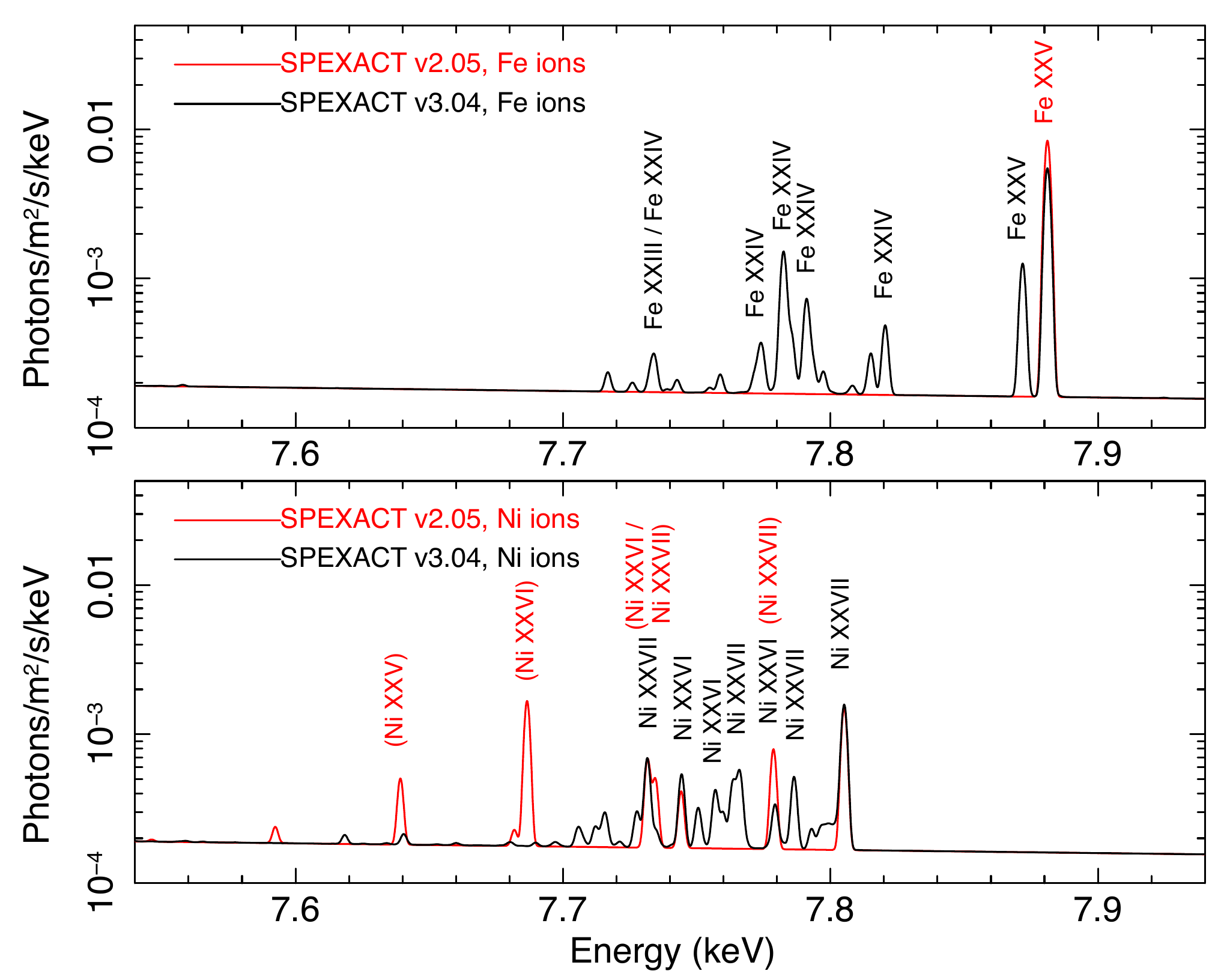} 
\caption{\textit{Left} \citep[from][reprinted with permission]{THC2017}: Same spectrum as Fig.~\ref{fig:Perseus_lines} right, zoomed on the Ni-K band. For comparison, the same spectrum observed by the \textit{XMM-Newton} EPIC instruments is shown in blue. \textit{Right} \citep[from][reprinted with permission]{mernier2018b}: Emission lines modelled for a $kT = 3$ keV CIE plasma using \textsc{SPEXACT} v2.05 (red) and \textsc{SPEXACT} v3.04 (black). For clarity, the Fe and Ni transitions are shown separately in the upper and lower panels, respectively.}
\label{fig:Ni_improvements}       
\end{figure}

\subsection{\textit{Hitomi} (and spectral model improvements)}
\label{sec:SN_models_Hitomi}

A breakthrough has been recently achieved with the {\it Hitomi} SXS observation of the core of the Perseus cluster. The left panel of Fig.~\ref{fig:Ni_improvements} presents the SXS spectrum in the 7.4--8.0 keV band, showing the He-like Ni resonance line clearly separated from the stronger Fe He$\beta$ line and other satellite emission. In addition, and compared to previous studies, the database \textsc{SPEXACT} used to fit this spectrum has undergone a major update, with the incorporation of 400 times more metal lines than in its previous version \citep[up to 2016; see e.g.][]{deplaa2017b}. Similar major improvements have been performed on the database AtomDB since the first release of the Perseus spectrum observed by \textit{Hitomi}, with no less than seven updates from then till now \citep[see e.g.][]{THC2018_atomic}.
These model updates -- in particular the \textsc{SPEXACT} ones -- have important consequences, especially around the 7.4--8.0 keV band where CCD instruments cannot disentangle the Ni-K lines from other particular Fe transitions (see the right panel of Fig.~\ref{fig:Ni_improvements}). Together, these two major improvements (spectral resolution and atomic model) allowed the most accurate measurement of the Ni/Fe abundance ratio in the ICM so far. As mentioned in Sect.~\ref{sec:metals_history}, the SXS also allowed the detection of weak resonance lines from Cr and Mn with high statistical significance. Flux measurements of these lines in individual objects have been more challenging to constrain with CCD detectors because such weak features easily blend into the continuum emission, yielding possible biases in their derived metal abundances. 

\begin{figure*}
  \includegraphics[width=1.0\textwidth]{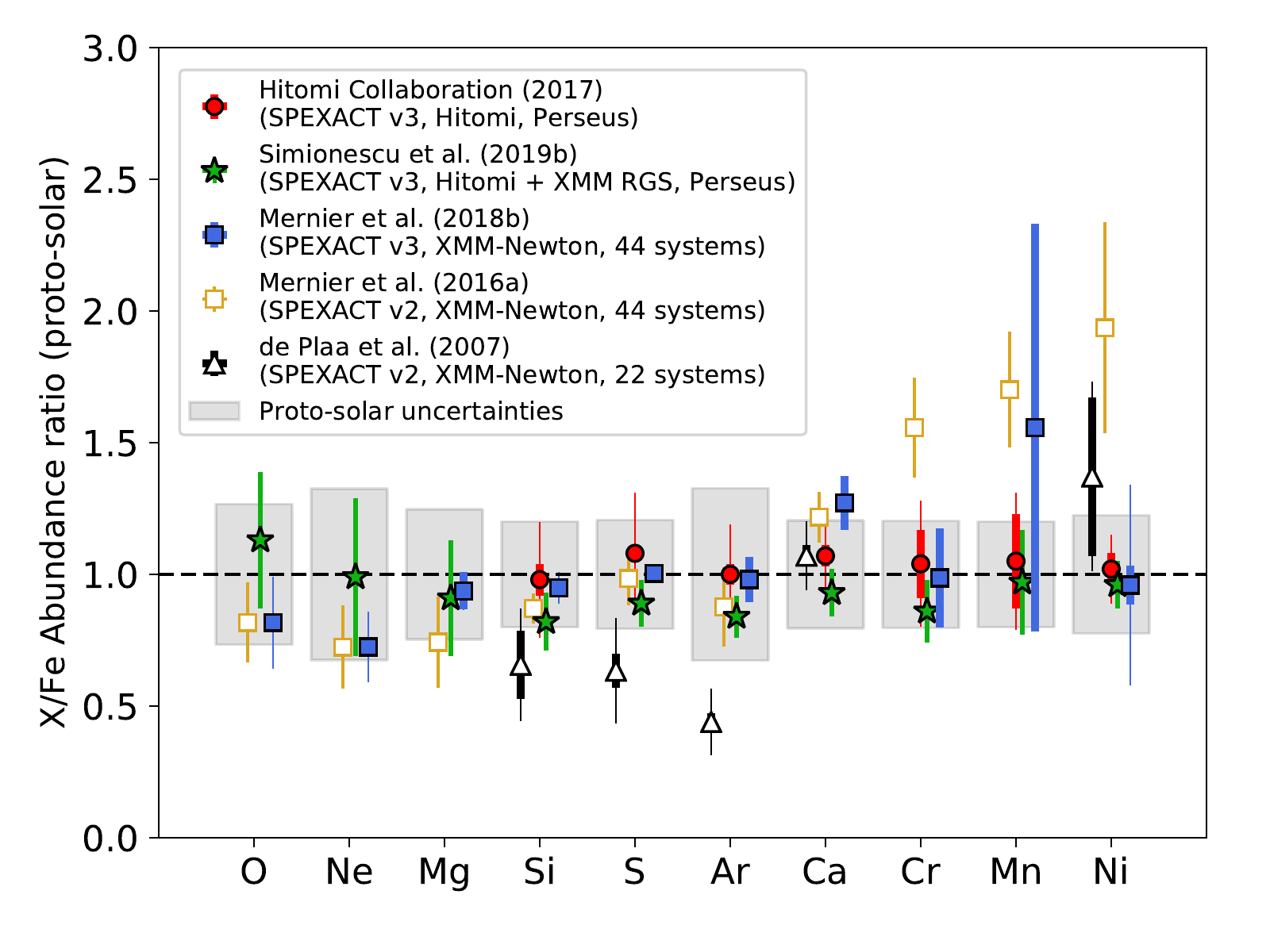}
\caption{Abundance ratios measured in the ICM compiled from the recent literature. All the ratios have been rescaled to the proto-solar values of \citet{lodders2009}. For comparison, uncertainties on the proto-solar values are shown by the grey filled rectangles.}
\label{fig:ICM_ratios}       
\end{figure*}

Figure~\ref{fig:ICM_ratios} compares the abundance ratios measured by several previous studies, including {\it Hitomi} \citep[][red circles]{THC2017} and {\it XMM-Newton} \citep[][black triangles and yellow squares, respectively]{deplaa2007,mernier2016a}. While the Fe-relative abundances of Si, S, Ar and Ca (whose K-shell emission lines are relatively strong) are consistent between the two most recent studies \citep{mernier2016a,THC2017}, the {\it Hitomi} measurement obtained significantly lower Cr/Fe, Mn/Fe and Ni/Fe ratios, revealing for the first time the ICM abundance pattern to be fully consistent with the solar composition. This new result suggests that near-Chandrasekhar-mass SNIa significantly contribute to the cosmic chemical evolution, at least in Perseus. 

In a follow-up work, \citet{simionescu2018} attempted to provide the most robust constraints of elemental ratios in the core of Perseus based on high-resolution spectroscopy, and re-evaluating the confidence ranges for the Si/Fe, S/Fe, Ar/Fe, Ca/Fe, Cr/Fe, Mn/Fe, and Ni/Fe ratios for \textit{Hitomi} to include uncertainties in the effective area calibration (Fig.~\ref{fig:ICM_ratios}, green stars). The abundance pattern is confirmed to be remarkably consistent with the Solar composition, but is challenging to reproduce with linear combinations of existing supernova nucleosynthesis calculations, particularly given the precise measurements of intermediate $\alpha$-elements enabled by Hitomi. Despite some degree of degeneracy that persists between several sets of models, \citet{simionescu2018} suggest that, under the reasonable assumption that SNIa and SNcc progenitors should have the same initial metallicity, including updates of neutrino physics in the core-collapse supernova yield calculations may improve the agreement with the observed pattern of $\alpha$-elements in the Perseus Cluster core. Further reduction of the current uncertainties on the measured ratios and on the predicted nucleosynthesis by stars and SNe will help to place more accurate constraints on the currently competing SNIa and SNcc models \citep[see also][for similar conclusions]{mernier2016b}. 

Moreover, and interestingly, \citet{simionescu2018} and \citet[][Fig.~\ref{fig:ICM_ratios}, blue squares]{mernier2018b} found that the previous tension between the {\it Hitomi} results \citep{THC2017} and the previous CCD measurements \citep{mernier2016a} can be largely alleviated when refitting {\it XMM-Newton} EPIC spectra with a consistent up-to-date version of \textsc{SPEXACT} (v3.04). This strongly suggests that despite their modest spectral resolution, CCD data are still able to constrain ICM abundances with acceptable reliability, as long as appropriate updates of atomic code are applied and systematic cross-calibration uncertainties are taken into account.

From Fig.~\ref{fig:ICM_ratios}, it is also remarkable to note that the level of accuracy reached by these updated (micro-calorimeter and CCD) measurements of the ICM is significantly greater than the current accuracy we have of the chemical composition of our own Solar System. Despite these accurate, converging, and encouraging measurements, it is useful to keep in mind that {\it Hitomi} was able to observe only one object within one specific spatial region. Better constraints of SN models by measuring the ICM abundances in a comprehensive way will require to study other clusters systematically with future high resolution spectroscopy instruments.

\subsection{Comparison with stellar abundances}
\label{sec:SN_stars}

\begin{figure*}
  \includegraphics[width=1.0\textwidth]{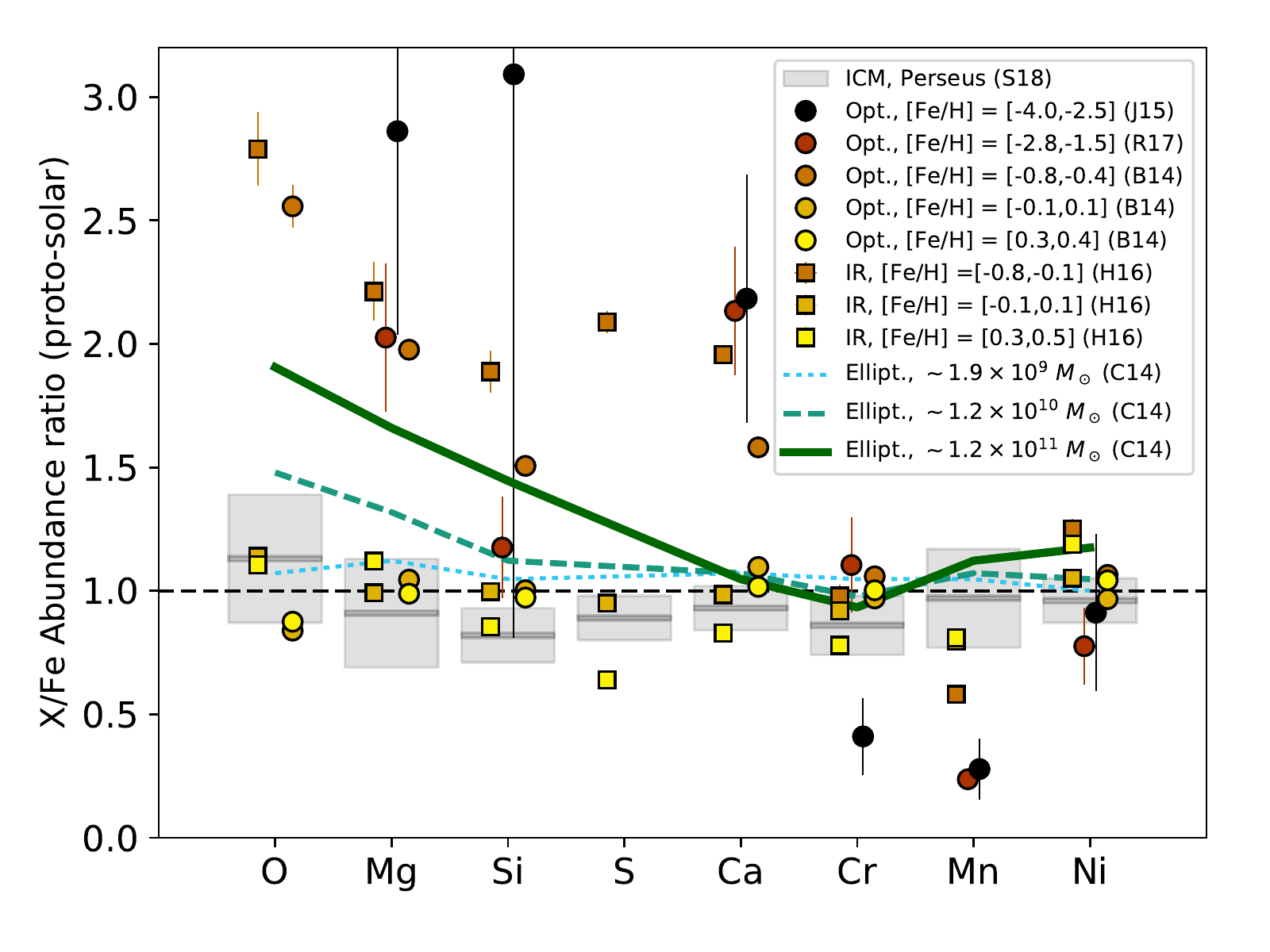}
\caption{Comparison between the abundances measured in the ICM of the Perseus cluster \citep[grey boxes;][]{simionescu2018}, stellar abundances in our Galaxy (circle and square data points), and stellar abundances in early-type galaxies \citep[solid, dashed, and dotted lines;][]{conroy2014}. Stellar Galactic abundances are estimated either via optical \citep{bensby2014,jacobson2015,reggiani2017} or infrared \citep{hawkins2016} observations. This figure is adapted from \citet{simionescu2018}.}
\label{fig:ICM_stars}       
\end{figure*}

Figure~\ref{fig:ICM_stars} compares the chemical composition of the ICM in Perseus with that of stellar populations in our Milky Way \citep[see also figure 6 in][]{simionescu2018}. As seen on the figure, metal-poor stars exhibit higher $\alpha$/Fe ratios \citep{jacobson2015,reggiani2017} while, on the contrary, stars with solar or super-solar metallicities show abundance patterns that are comparable to that of the Sun and of the ICM \citep{bensby2014,hawkins2016}. This trend is expected, because stellar metallicity correlates with age \citep{twarog1980} and older stars naturally incorporate less SNIa products than younger ones. For a complete review on recent stellar abundance measurements (and the Galactic chemical enrichment), we refer the reader to \citet{nomoto2013}.

Whereas a typical late-type galaxy like our Milky Way experiences multiple episodes of star formation even at present times, galaxies in cluster and groups -- in particular their central BCG -- are mainly early-type with a star forming activity that has considerably quenched \citep[see also][in this topical collection]{werner2018}. At first glance, the chemical composition of the ICM should thus be better compared to the stellar populations of these "red-and-dead" galaxies. As shown in Fig. \ref{fig:ICM_stars} (solid, dashed, and dotted lines), the bulk of stars in massive ellipticals exhibit super-solar $\alpha$/Fe ratios \citep{thomas2005,conroy2014}. The main interpretation is that star formation in the most massive ellipticals is shorter, hence SNIa explode too late for their products to be efficiently incorporated into these stars. On the other hand, very few SNcc are found in ellipticals at present times \citep{graham2012} and products of recent enrichment in the interstellar medium of such galaxies are thought to be dominated by SNIa \citep{mannucci2008}. For this reason, the remarkable similarity of the chemical composition of the ICM (and of ellipticals hot atmospheres -- see Sect. \ref{subsec:budget_composition}) with that of the Solar neighbourhood is surprising and not trivial to understand. This question becomes even more difficult when considering that the bulk of the ICM enrichment likely occurred at $z \gtrsim $ 2--3 (Sects.~\ref{sec:distribution} and \ref{sec:redshift}), i.e. when the relative SNcc-to-SNIa contribution in the Universe was likely more important than today. A debatable -- though not fully excluded -- possibility would be that enrichment by stellar mass loss and by SNIa explosions compensate each other exactly to recover the near-solar chemical composition measured in the core of Perseus \citep{THC2017,simionescu2018} and most other relaxed systems \citep{mernier2018a}. Alternatively, the relative contribution of late time SNIa explosions to the enrichment could be negligible (see Sect. \ref{sec:radial_core}).

\subsection{Nitrogen and enrichment from AGB stars}
\label{sec:nitrogen}

Using the RGS instruments on board \textit{XMM-Newton}, a couple of studies also detected significant signatures of N in the hot atmospheres of giant ellipticals \citep{xu2002,tamura2003,werner2006b,sanders2008,sanders2010,grange2011,sanders2011}. To date, the most complete study on N abundance in groups and ellipticals is presented in \citet{mao2019}, where the authors reported significantly higher N/O and N/Fe abundance ratios than those measured in various stellar populations of the Galaxy. Their results also suggest that the bulk of the N enrichment seen in these hot haloes predominantly originates from AGB stars rather than SNcc (see also Sect.~\ref{sec:AGB}). Because the RGS is a grating spectrometer, however, metal lines are broadened by the spatial extent of the source. For this reason, detecting weak lines such as N in more extended sources than compact ellipticals (e.g. clusters) remains very challenging with the current instruments.


\section{Chemical evolution with redshift}
\label{sec:redshift}

\begin{figure*}
  \includegraphics[width=1.0\textwidth]{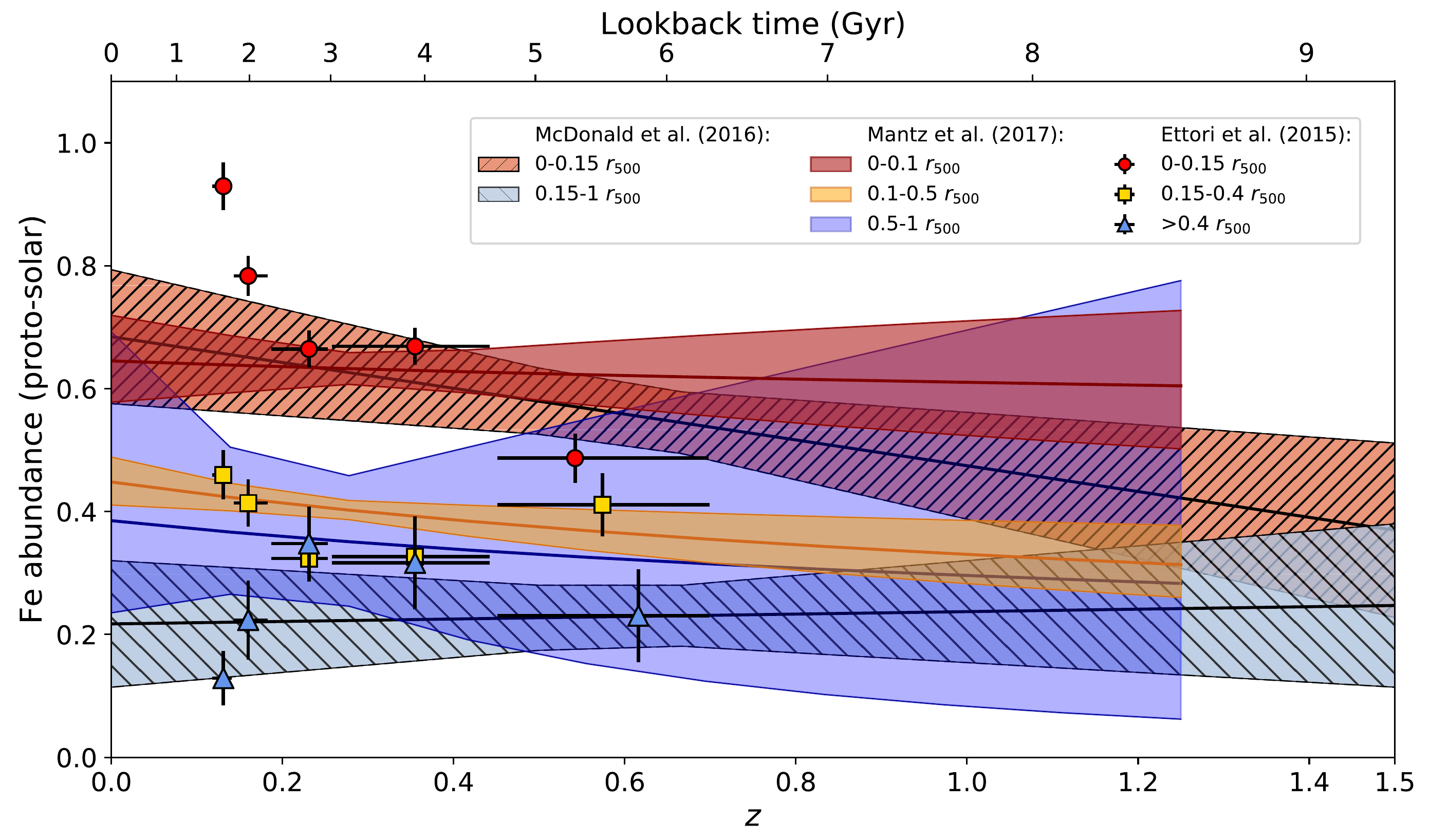}
\caption{Redshift evolution of the Fe abundance within different radial bins in cool-core clusters, compiled from recent works.  For clarity, the scatter of the measurements are not shown. All the abundances are rescaled with respect to the proto-solar values of \citet{lodders2009}.}
\label{fig:redshift_Fe}       
\end{figure*}

Since the first attempt to measure the average abundance of Fe in 18 hot galaxy clusters out to redshift $\sim 1.3$ \citep{tozzi2003}, 
several studies \citep{balestra2007,maughan2008,anderson2009,andreon2012,baldi2012,ettori2015,mcdonald2016,mantz2017} have tried to assess if, and to quantify by how much, the ICM metallicity evolves both in the core and outer regions.
We report here on the most recent works that have addressed this issue both increasing the sample of spatially resolved spectral measurements and the accuracy in the sample selection. The redshift-metallicity trend for each of these studies is illustrated in Fig.~\ref{fig:redshift_Fe}.

\cite{ettori2015} considered a hetereogenous sample of 83 objects with sufficient signal in {\it XMM-Newton} exposures to resolve the metallicity measurements in three radial bins ($0$--$0.15$, $0.15$--$0.4$ and $>$0.4$\,r_{500}$) out to redshift 1.4. They showed that any significant evidence ($>$3$\sigma$) of evolution of the metallicity with redshift is limited to the inner regions of cool-core clusters.
\cite{mcdonald2016} selected about 150 clusters observed with \textit{Chandra}, \textit{XMM-Newton} and \textit{Suzaku} from a mass-selected SPT (South Pole Telescope) sample and concluded that, since $z=1$, the metallicity is changed by no more than 40\%, strongly suggesting an early ($z>1$) enrichment. Evidence for evolution in the central region of CC clusters was found, although with weaker significance than previously reported.
\cite{mantz2017} used \textit{Chandra} data of 245 clusters selected from X-ray and Sunyaev-Zel'dovich (SZ) surveys up to $z=1.2$ and measured a moderate late-time increase at intermediate radii. This, combined with recent hints that central Fe peaks are on average sharper at higher redshift \citep{liu2018}, is consistent with ongoing mixing of metals in central regions, possibly out to intermediate radii, through gas sloshing and AGN feedback (see Sect.~\ref{sec:inhomogeneities}).

Altogether, these results tend to support the current picture of the early enrichment scenario as introduced in Sect.~\ref{sec:radial_outskirts}, in which the metallicity in the outskirts did not evolve in the last $\sim$6--9 Gyr. Observational results also point towards a weak evolution of central metallicities with time, although conclusive evidence is still lacking. On the other hand, the Fe abundance found in excess with respect to the mean ambient value in the WARP-J1415 cluster \citep[][]{degrandi2014} suggests that even in cluster cores, most of the enrichment was already in place at $z\simeq 1$ (see also Sect.~\ref{sec:radial_core}).

It is quite remarkable to note that, following our current understanding of the early enrichment scenario, the presence of metals at relatively high-$z$ in the inter-galactic medium could be only explained from a theoretical perspective if early AGN feedback is taken into account \citep[as stellar feedback alone is not efficient enough to eject metals outside galaxies;][in this topical collection, and references therein]{biffi_review}. In other words, probing the (absence of) metallicity evolution from the ICM assembling epoch up to now is also an indirect diagnostics of supermassive black hole activity in the early Universe.

One of the Science Goals for the next generation of the X-ray telescopes such as {\it Athena} is, therefore, to provide more robust measurements of Fe and especially $\alpha$/Fe at high-$z$ in order to further constrain the enrichment history over cosmic time \citep[][see also Sect.~\ref{sec:conclusion}]{pointecouteau2013,cucchetti2018}.


\section{Metal budget in galaxy clusters, groups, and elliptical galaxies}
\label{sec:budget}

\subsection{Is the ICM too enriched?}
\label{subsec:budget_clusters}

Whereas a significant fraction of metals are ejected from supernovae and escape beyond their host galaxies, the remaining part remains locked in the local ISM and directly contributes to the formation of the next generation of stars. Quite surprisingly, comparisons between stellar light and ICM abundances suggest that there are at least as much metals dispersed in elliptical/group/cluster hot atmospheres than locked in stars \citep[e.g.][]{renzini1993}. Quantitatively, reconciling such a large amount of intra-cluster metals -- in particular Fe in massive clusters -- with what stars in galaxies could have reasonably produced has constituted a serious challenge for several decades \citep[e.g.][]{vigroux1977,arnaud1992,loewenstein2006,bregman2010,loewenstein2013,renzini2014}. 

Several possibilities were proposed to solve this metal budget paradox in clusters. Among them, an incorrectly assumed IMF in clusters -- being rather "top heavy", hence boosting the generation of massive stars -- could help to synthesize and release more Fe in the ICM via SNcc explosions \citep[e.g.][]{portinari2004,nagashima2005}. The existence of such a "top-heavy" IMF, however, remains challenging to confirm or rule out via the ICM abundance pattern because of the still large error bars of the abundance ratios of interest \citep[including O, Ne, and Mg;][see also Fig.~\ref{fig:SN_models} \textit{top} and Fig. \ref{fig:ICM_ratios}]{mernier2016b}. Moreover, \citet{matsushita2013} found that a top-heavy IMF is disfavoured in the Perseus cluster because it would overproduce 
the Si mass-to-light ratio (and presumably that of other $\alpha$-elements). Finally, evidence is accumulating towards a "bottom-heavy" IMF in early-type galaxies \citep[e.g][]{vandokkum2010,conroy2012,goudfrooij2013}, thereby worsening the paradox. 

Several alternative scenarios are also considered, among which underestimated efficiencies for metals to be released into the ICM \citep[however, see][and references therein]{loewenstein2013}, higher SNIa rates with redshift \citep[][see however Sharon et al. \citeyear{sharon2010}]{perrett2012} and/or in cluster environments \citep{maoz2017,friedmann2018}, a significant contribution of the ICM enrichment by Population III stars and/or pair-instability SNe \citep[e.g.][]{morsony2014}, or an enriching stellar population which is distinct from cluster galaxies \citep[in particular the intracluster stars; e.g.][]{zaritsky2004,gonzalez2007,bregman2010}. An interesting alternative may be proposed again in the context of an early enrichment scenario. If this is the case, only the lowest mass stars (with a lifetime of several Gyrs, i.e. compatible with that early enrichment epoch) seen in cluster galaxies would be related to the metals seen in the ICM today. However, they would not be the dominant stellar population seen in optical images, and today's stars would have little to do with the bulk of the ICM enrichment. Qualitatively, this scenario could explain this paradox while remaining in line with the early enrichment scenario discussed above. Quantitatively, however, this needs to be demonstrated via deeper stellar and/or ICM observations and numerical simulations.

\subsection{Are ellipticals and galaxy groups less enriched than clusters?}
\label{subsec:budget_groups}

Metallicities in the central ICM regions of massive ellipticals, groups, and clusters are often measured individually or in small samples belonging to one of these three categories. Few studies, surprisingly, attempted to compare Fe abundances (or abundances of any other element) within consistent regions for all these different systems. Until recently, the main picture was that, within $\sim$$r_{500}$ or less (depending on the studies), groups and ellipticals are significantly less Fe-rich than clusters, with a positive correlation between the Fe abundance and temperature (or mass) of less massive systems, whereas this dependence became weaker in the cluster regime \citep{fukazawa1998,rasmussen2007,rasmussen2009,sun2012,mernier2016a,yates2017}.

Such a difference of metallicity between hot, massive clusters and cool, less massive systems was not easy to explain. Instead, theoretical predictions, e.g. from simulations or semi-analytic models, show a very weak dependence of the metallicity on the system mass -- not only in clusters but also in groups and ellipticals, with a larger scatter at low masses \citep[][in this topical collection, and references therein]{biffi_review}. Tentative scenarios to explain this tension between observations and simulations included for instance a depletion of metal-rich material out of the X-ray phase (e.g. cold filaments), selective removal of metals via powerful AGN feedback in shallower gravitational wells, or even different star formation histories between clusters and less massive systems \citep{rasmussen2009}.

\begin{figure}
\begin{center}
  \includegraphics[width=0.8\textwidth]{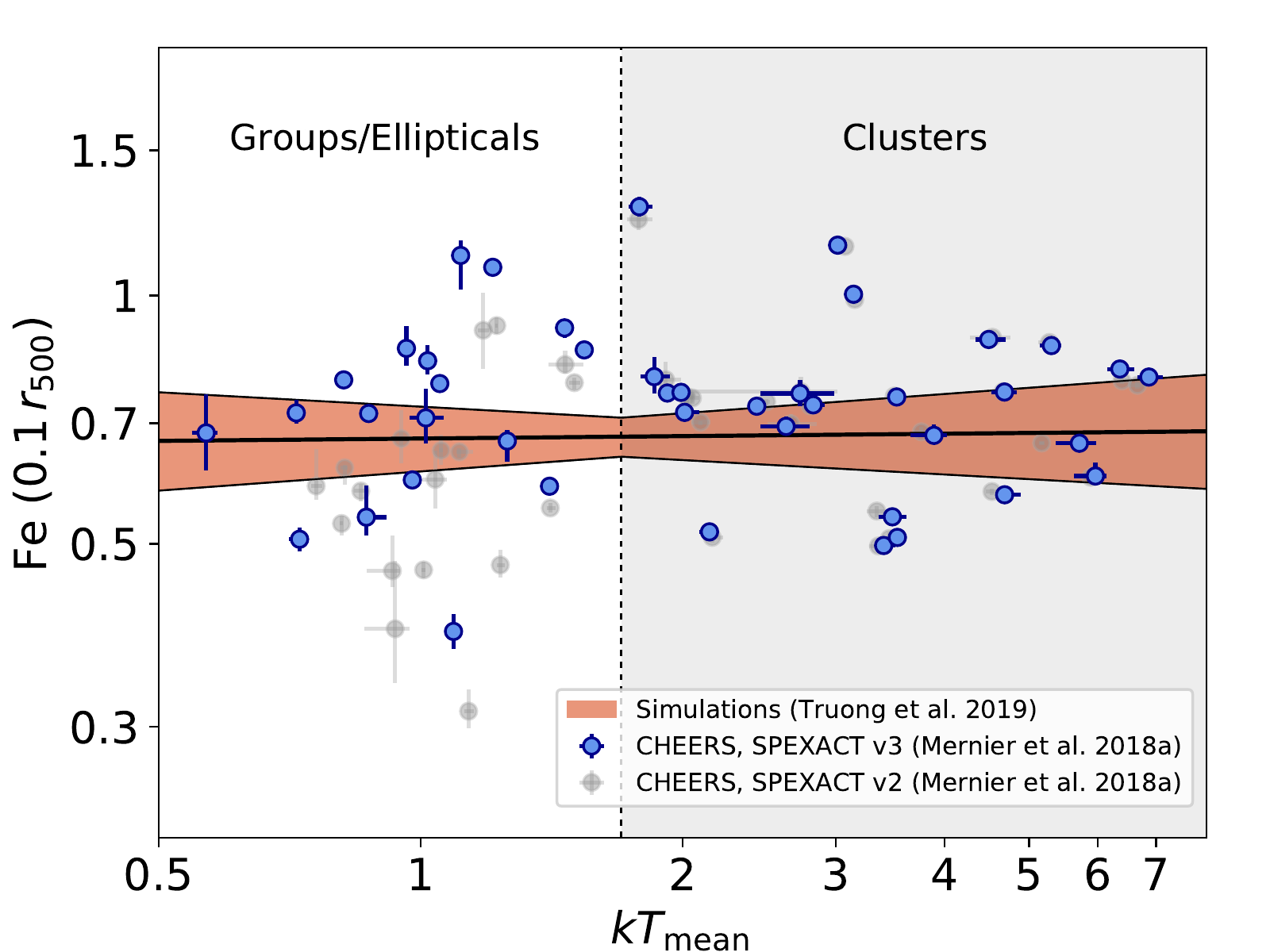} \\
  \includegraphics[width=0.9\textwidth]{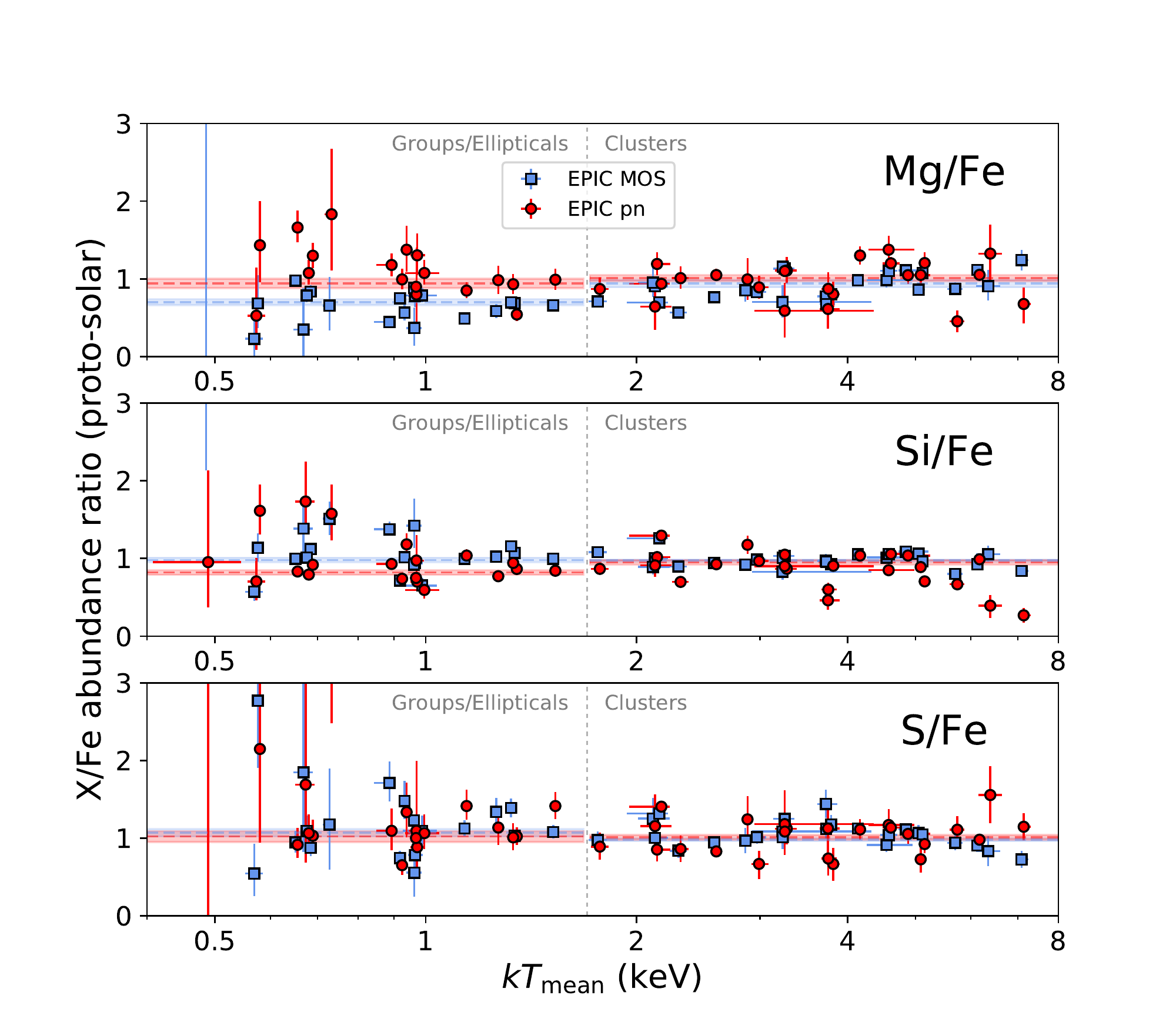} 
\end{center}
\caption{\textit{Top}: Observations (CHEERS sample) vs. simulations of central ($\le 0.1 \,r_{500}$) Fe abundances in the ICM of systems including ellipticals, groups, and clusters. The filled area corresponds to the 68\% confidence region of the simulated systems \citep{truong2019}. The effects of recent updates of the spectral code \textsc{SPEXACT} (v3) are seen in the observed abundances of low-temperature systems \citep{mernier2018a}. \textit{Bottom} \citep[from][reprinted with permission]{mernier2018b}: Mg/Fe (upper panel), Si/Fe (middle panel), and S/Fe (lower panel) abundance ratios in the central ICM of ellipticals, groups, and clusters (CHEERS sample).}
\label{fig:cl_gr}       
\end{figure}

On the other hand, it should be kept in mind that the Fe abundance measurements in clusters and groups/ellipticals are based on very different lines and energy bands. While most of the Fe-K transitions (used to constrain Fe in clusters) are now well understood \citep[e.g.][]{THC2018_atomic}, the Fe-L complex (used to constrain Fe in groups and ellipticals) is unresolved by CCD instruments and may easily bias the Fe abundance measurement in many ways (Sect.~\ref{sec:biases}). In particular, a major update of the spectral atomic code \textsc{SPEXACT}, incorporating more than one million of new transitions (see also Sect.~\ref{sec:SN_models_Hitomi}), changes the (still unresolved) modelled shape of the Fe-L complex in cool plasmas. As a consequence (Fig.~\ref{fig:cl_gr}, \textit{top}), the Fe abundance of groups and ellipticals within 0.1$\,r_{500}$ has been recently revised upwards and measured to be very similar to those in clusters \citep{mernier2018a}, thereby providing a much simpler picture on this question \citep{truong2019}. 

The situation is less clear for isolated massive spiral/lenticular (S0) galaxies. Among the rare studies of the hot atmospheres surrounding S0s, metallicities have been constrained in NGC\,1961 \citep{anderson2016} and NGC\,6753 \citep{bogdan2017}, both with rather low central values ($\sim$0.1--0.3 solar) compared to more massive systems. In these two cases, however, the authors caution against systematic biases (e.g. the Fe-bias, Sect. \ref{sec:Fe_bias}) that may significantly affect their measurements. In fact, when assuming a multi-temperature distribution, \citet{juranova2019} find a metallicity of $\sim$0.7 Solar for NGC\,7049, consistent with the average value of more massive ellipticals, groups, and clusters. More studies of S0s, however, are required to obtain a better picture of their typical hot gas-phase enrichment.

Further exploration, confirmation, and limits of the mass-invariance of gas metallicity are expected with better spectral resolution instruments onboard future missions (e.g. \textit{XRISM}, \textit{Athena}; see Sect.~\ref{sec:conclusion}). In particular, extending these comparisons to the outskirts of all these systems will be of high importance.

\subsection{Origin and chemical composition of ellipticals' hot atmospheres}
\label{subsec:budget_composition}

Thanks to the state-of-the-art cosmological simulations, the primordial origin of the ICM of the largest scale structures -- i.e. accreting baryons getting rapidly thermalised via shocks and enriched early in their history -- starts to be well established \citep[see][in this topical collection]{biffi_review}. The origin of the hot atmospheres surrounding isolated elliptical galaxies, however, is less obvious. The reason is that at these smaller scales (less easily reachable by cosmological simulations) and shallower gravitational wells, stellar and AGN feedback might start to play a significant role. Specifically, the relative contribution of infalling gas vs. stellar mass loss giving rise to the observed hot atmospheres in ellipticals is not well known. Whereas the origin of this (X-ray emitting) interstellar medium had long been attributed mainly to stellar mass loss \citep[e.g.][]{mathews1990,ciotti1991,sarzi2013}, the current emerging picture instead suggests that early infall is of first importance \citep[e.g.][Werner et al. \citeyear{werner2018}, in this topical collection]{goulding2016,forbes2017}.

Interestingly, the observed chemical composition of these hot atmospheres can also help to probe their origin. Although, like for absolute metallicities (Sect.~\ref{subsec:budget_groups}), very few comparisons have been established between the chemical composition of ellipticals and that of groups/clusters, \citet{mernier2018b} found that X/Fe abundance ratios are very close to solar for each of the CHEERS systems, independently on their mass (Fig.~\ref{fig:cl_gr}, bottom). In line with the recent results reported above, this remarkable similarity in the chemical composition of ellipticals, groups, and clusters also suggests a similar origin and formation of the hot gas in all these systems. A similar approach could be extended to massive spiral/lenticular (S0) galaxies. However the weak X-ray luminosity of their hot interstellar medium requires very deep exposures, and to our our knowledge, only absolute metallicity measurements have been reported so far \citep[][Sect.~\ref{subsec:budget_groups}]{anderson2016,bogdan2017,juranova2019}. 

As illustrated here and in Sect. \ref{subsec:budget_groups}, the extension of chemical enrichment studies to hot atmospheres of lower mass systems will undeniably be a topic of higher interest in the near future.


\section{Current uncertainties and biases in measuring abundances}
\label{sec:biases}

Although deriving and constraining elemental abundances in the ICM is, in principle, relatively simple (Sect.~\ref{sec:lines_abundances}), in practice such estimates may be affected by several systematic effects and biases. In this section we enumerate a (non-exhaustive) list of known effects which may significantly bias abundance measurements that have been reported in the literature so far.

\subsection{Effect of possible He sedimentation on the other abundance measurements}
\label{sec:sedimentation_biases}

As was mentioned in Sect.~\ref{sec:lines_abundances}, the spectroscopic inference of elemental abundances is essentially based on a comparison of the strength of metal lines with the underlying continuum.  Implicitly, this procedure involves an assumption about abundances for the elements that contribute significantly to the continuum. While an incorrect line modelling (due to, e.g., atomic data uncertainties, see Sect. \ref{sec:atomic_biases}) may evidently result in incorrect abundance estimates, an incorrectly calculated continuum level caused by a non-solar abundance distribution may equally affect the corresponding equivalent width, hence the absolute abundance of the investigated element. Here we discuss this latter source of uncertainties.

At typical X-ray temperatures the free-free continuum is essentially dominated by thermal bremsstrahlung (Sect.~\ref{sec:lines_abundances}).
For a fully ionized plasma with solar composition,  H and He contribute to the X-ray free-free continuum with a ratio of about 2:1, while the contributions from other elements are much smaller. 
At lower temperatures, however, bound-free processes are also important, 
in which emission from C, N and O is noticeable (besides H and He).
As seen from Eq.~\ref{eq:cont},  the thermal bremsstrahlung emissivity depends on the He abundance as follows:
\begin{equation}
\epsilon_{ff} \propto n_e (n_p+Z^2_{{\rm He}}n_{{\rm He}}) \approx n_p^2 (1 + 2x) (1 + 4x),
\end{equation}
where $n_p$, $n_{{\rm He}}$, $n_e$  are the plasma H, He and electron densities and $x= n_{{\rm He}}/n_p$ is the He abundance.

It is well known that the He abundance in the ICM can not be directly measured by spectroscopic means, because both H and He are fully ionized at ICM temperatures and produce no spectral lines.
For this reason, the He abundance is unknown and, in practice, is usually assumed to be equal to its primordial value (or to the proto-solar value, which is $4$ \% smaller for the set of \citealt{lodders2009}).
According to the big bang nucleosynthesis (BBN) theory, the majority of He, along with hydrogen and a small admixture of other light nuclides, were produced during the first few minutes of the Universe \citep[see][for a review]{coc2017}. 
Recently, the predictions of the standard BBN model for primordial plasma composition have reached unprecedented level of precision, thanks to anisotropy measurements of the cosmic microwave background (CMB) by the Planck space observatory \citep{planck2015_13}.
 Now the primordial He mass fraction is constrained to within a tenth of a per cent: 
 $Y^{\rm BBN}_{\rm p} = 0.2484 \pm 0.0002$ \citep{coc2017} (which corresponds to $x=0.0827$), 
 that is in general agreement with direct measurements \citep{izotov2014,aver2015}.
 Since that primordial epoch, the primordial plasma composition in most places of the Universe over time has been modified by stellar evolution.
However, the effect of the latter on the He abundance in the ICM is rather small, bearing in mind that the mass of primordial He is comparable to the total stellar mass in a cluster.

Significantly larger changes in the He abundance may be caused by sedimentation of 
elements heavier than hydrogen in a cluster's gravitational potential. 
A straightforward method to estimate the He sedimentation amplitude is based on solving Burgers'  equations \citep{burgers1969} for a 1D cluster model with no magnetic fields  \citep{fabian1977,gilfanov1984,ettori2006,shtykovskiy2010,medvedev2014}.
Based on diffusion calculations  with such a model  for a set of parameters of  cool-core clusters from \cite{vikhlinin2006}, the average He abundance inside $r_{2500}$ ($\simeq 0.4\,r_{500}$) grows at a rate of about 3\% in 1 Gyr at the same radius for a cluster of $10^{14}\,M_{\odot}$ of total mass \citep{medvedev2017}. The effect increases almost linearly with cluster mass and time. 
The change in the He abundance becomes more prominent at smaller radii. In the cluster inner core ($\lesssim 0.05\,r_{500}$), however, the efficiency of He sedimentation can be suppressed by thermal diffusion. Instead, He may peak at the intermediate radii $\sim$0.1--0.2$\,r_{500}$, where its abundance enhancement can reach 20\% after 1 Gyr of the diffusion  \citep{medvedev2014}.
At the same time, sedimentation can be inhibited by several mechanisms.
As is well known, transport processes in the ICM may be suppressed by tangled magnetic fields \citep{chandran1998,narayan2001}. 
The effects of the latter are usually parametrized by a constant suppression factor of diffusion coefficients \citep{chuzhoy2004, peng2009}. 
However, this approach  does not allow to investigate a wide variety of instabilities that can play an important role in the plasma dynamics \citep[see][]{berlok2015,berlok2016}.

Since for a CIE plasma the most important excitation process is collisions with free electrons, the line emissivity of an element $X$ is $\propto n_e n_X \sim (1+2x) n_p n_X$ (see Eq.~\ref{eq:line}). 
Then fixing the observable line-to-continuum ratio, the dependence of the derived abundance $n_X/n_p$ 
on the assumed helium abundance can be  estimated as: 
\begin{equation}
n_X/n_p \propto (1+4x).
\end{equation}
It implies that an underestimation  of the He abundance in a model fitting will mimic an underabundance of metals. For instance, 
an error of -50\% in the assumed He abundance results in $\sim$10\% bias in metal abundances. 
Finally, we note that He sedimentation can similarly affect various cosmologically important measurements with galaxy clusters,
such as determinations of the gas mass, total mass and angular distance  \citep{ettori2006, markevitch2007a, peng2009, medvedev2014}.

\subsection{Atomic code uncertainties}
\label{sec:atomic_biases}

The atomic codes that are being used for the analysis of X-ray spectra and the derivation
of the abundances in the hot gas show significant differences and have considerable uncertainties.
This was shown in detail for the \textit{Hitomi} spectrum of the Perseus cluster \citep{THC2018_atomic}. 

While the latest versions of the two main plasma codes, \textsc{APEC} and \textsc{SPEXACT}, were updated to deal with the high spectral quality of the \textit{Hitomi} data (see also Sect.~\ref{sec:SN_models_Hitomi}), they needed substantial improvements to obtain the best results. In retrospect, most of these updates had to do with outdated atomic data or software bugs, which however are hard to find systematically due to the large amount of calculations that is involved.

But even the updated codes result in some relatively subtle but noticeable differences. Starting with the element with the strongest lines, namely Fe, it was found that the Fe abundance as derived from the \textsc{APEC} code is 15\% lower compared with \textsc{SPEXACT}. The main reason is the use of different data sets for collisional excitation, and at this moment it is hard to say what the true excitation rate should be. Moreover, the Fe abundance is not determined from one line, but from several lines, some of which also suffer from other astrophysical effects like resonant scattering, that needs to be taken into account properly. Without accounting for resonant scattering \citep[e.g.][]{THC2018_rs}, the derived Fe abundance in the core of Perseus would be 11\% lower.

For the other elements measured by \textit{Hitomi} (Si, S, Ar, Ca, Cr, Mn and Ni), the differences for the derived abundances when using both codes typically range between 5--10\%, again due to different collisional excitation rates. Like Fe, which shows significant code-related discrepancies, it should be noted that the differences for these other elements are in general larger than the statistical uncertainties on their abundances.

While the abundances of the Perseus cluster core with \textit{Hitomi} were derived from the relatively simple K-complexes of the relevant elements, measurements with other satellites (either past, operational or future) also incorporate the Fe-L band. It is well known that there are several
uncertainties in predicted line powers for the strongest lines in that wavelength region, which may be up to tens of percents. It is unclear how much that will affect precisely the derived abundances, but preliminary studies \citep[][see also Sect.~\ref{subsec:budget_groups}]{mernier2018a} show that the effects can be large and affect our interpretations on the ICM enrichment.

Clearly, stronger and more focused efforts to improve the atomic data for plasma codes in this regime are required.

\subsection{The Fe-bias and inverse Fe-bias}
\label{sec:Fe_bias}

\begin{figure}
\begin{center}
  \includegraphics[width=0.8\textwidth]{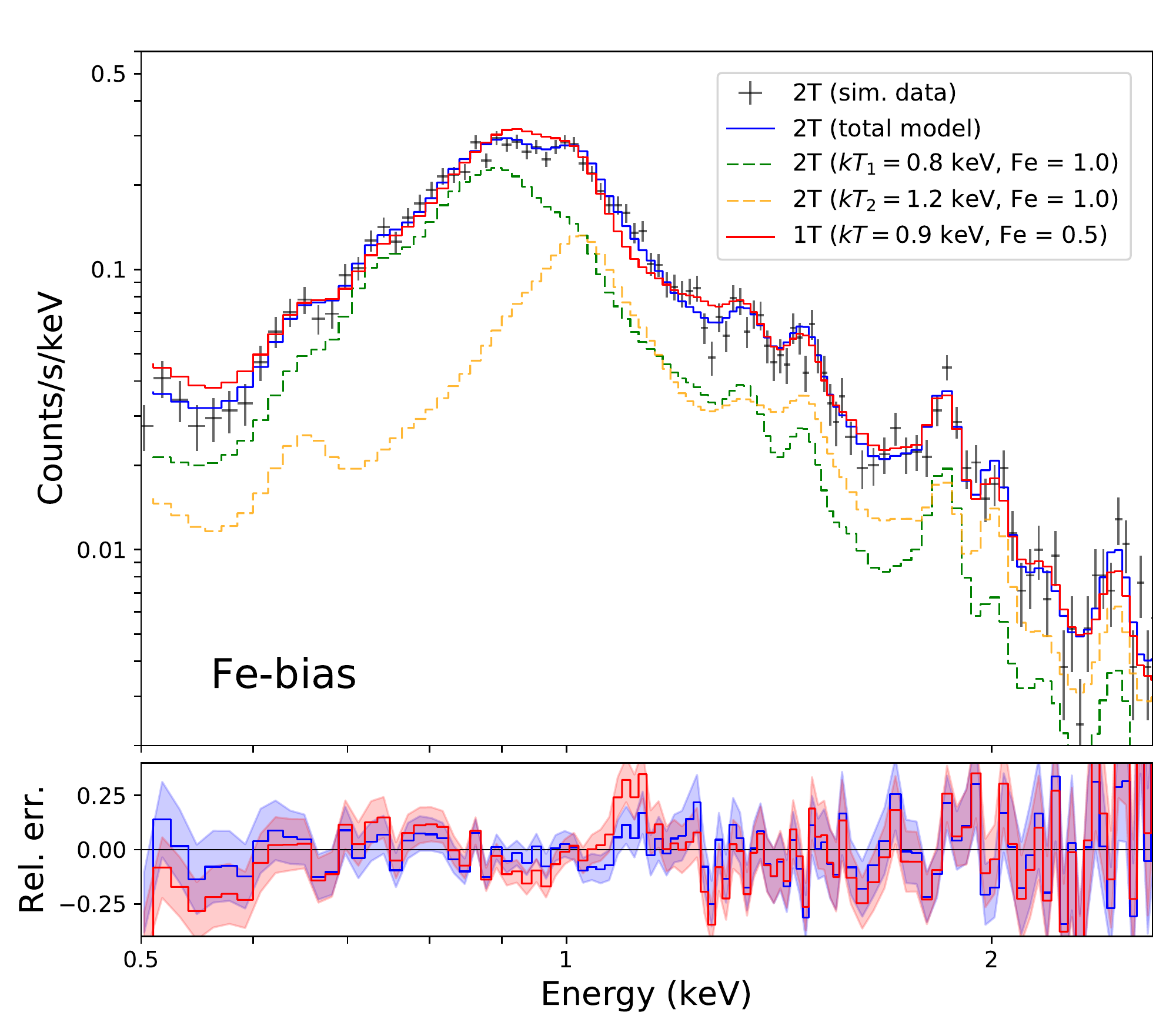} \\
  \includegraphics[width=0.8\textwidth]{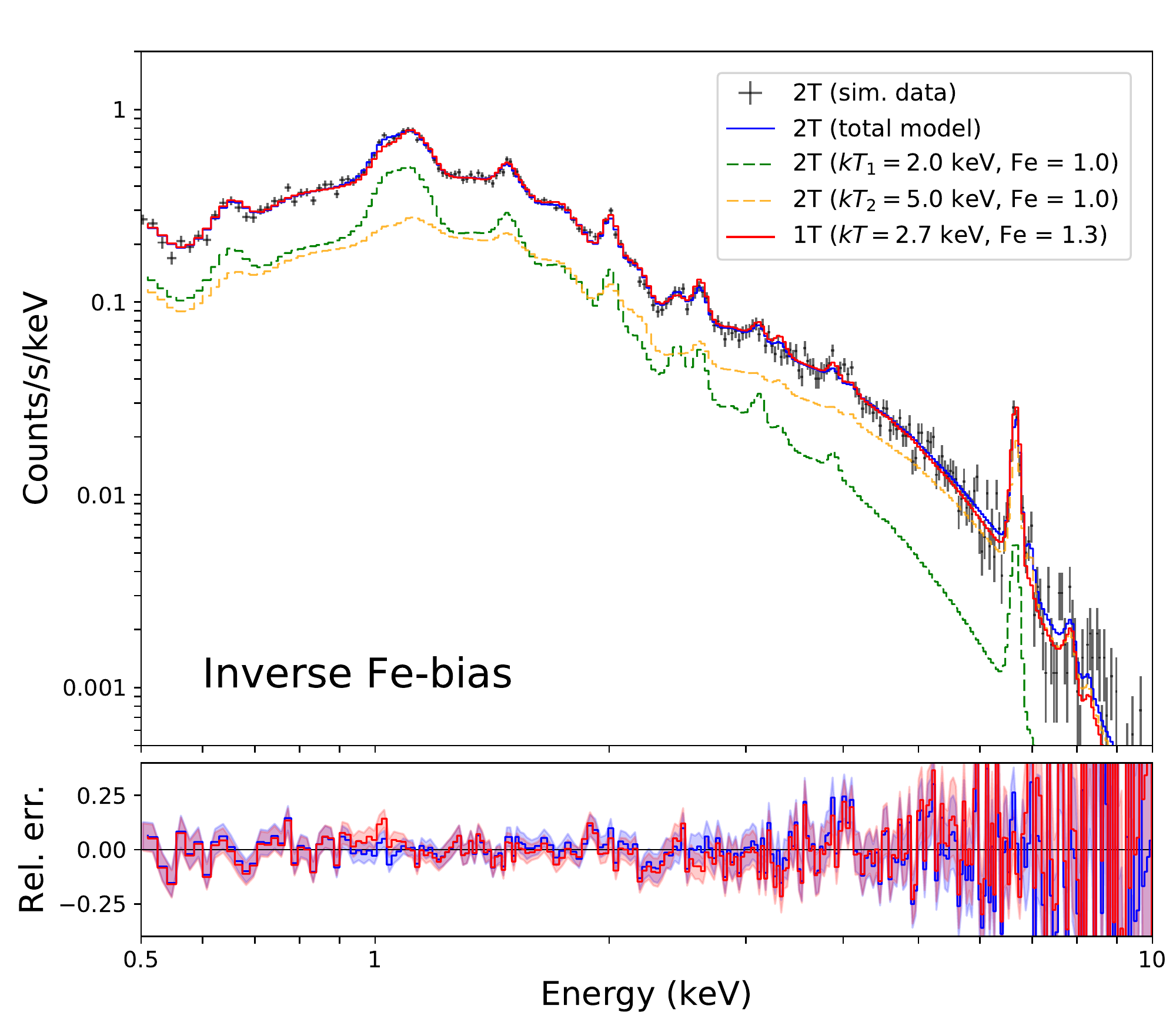} 
\end{center}
\caption{Illustration of the Fe-bias (\textit{top}) and the inverse Fe-bias (\textit{bottom}). In both bases, a two-temperature plasma (2T; with temperatures $kT_1$ and $kT_2$) is simulated through an EPIC MOS spectrum, which is then re-fitted using a single-temperature component (1T). In both cases, the 2T and 1T fits (blue and red lines, respectively) provide very similar spectra, yet with different Fe abundances.}
\label{fig:Fe_bias}       
\end{figure}

The Fe-bias tends to underestimate significantly the Fe abundance in galaxy groups and ellipticals if only one single-temperature plasma is assumed \citep{buote1994,buote1998,buote2000}. Given that the (mostly central) ICM region under investigation is never isothermal (because of projection effects on temperature gradients and a possible instrinsic multi-phaseness of the gas), this bias is easy to understand (Fig.~\ref{fig:Fe_bias}, \textit{top}). Assuming for simplicity that the gas in a galaxy group is actually made of two single-temperature components (say, $kT_1 \simeq 0.8$ keV and $kT_2 \simeq 1.2$ keV), the cooler (hotter) component excites in priority lower (higher) energy transitions of the Fe-L complex. The overall shape of the Fe-L complex, combining these lower and higher components, will thus appear flatter, more extended, and centred on transitions of intermediate energies. Such an overall spectral shape will be naturally reproduced by an isothermal model of intermediate temperature (roughly 0.9 keV) and with lower Fe abundance.

Like the Fe-bias, the inverse Fe-bias is also due to the (overly simplistic) assumption of a single-temperature gas component to model the ICM. Compared to the former, however, the inverse Fe-bias concerns hotter systems (with, typically, $kT \simeq$ 2--4 keV) in which both the Fe-L complex and the Fe-K transitions contribute equally to the Fe estimation, and results in an overestimation of the Fe abundance \citep{rasia2008,simionescu2009,gastaldello2010}. This bias is also easy to understand in the case of a two-temperature plasma (with, say, $kT_1 \simeq 2$ keV and $kT_2 \simeq 5$ keV) modelled with one single-temperature component (Fig.~\ref{fig:Fe_bias}, \textit{bottom}). While in the lower-temperature component the Fe abundance is reflected mainly through the Fe-L complex, a similar Fe abundance is observed in the high-temperature component via the Fe-K transitions. Consequently, the only way to reproduce simultaneously the boosted emissivities of these two spectral features with one single-temperature component is to (incorrectly) increase the best-fit Fe abundance parameter.

\subsection{Other biases and uncertainties}
\label{sec:other_biases}

Among the potential other biases and sources of uncertainty affecting the current measurements, one can also mention the following items.

(i) Although the instrumental calibration of \textit{XMM-Newton}, \textit{Chandra}, and \textit{Suzaku} has considerably improved with years, it should be kept in mind that no instrument is perfectly calibrated or understood. Cross-calibration mismatches are known to affect the measured temperature of clusters \citep{schellenberger2015}, which in turn may affect abundance measurements. In addition, imperfections of the effective area may locally under- or overestimate the continuum around specific metal lines. Since the flux of the continuum is directly used to calculate the equivalent width of the lines, this can substantially bias the inferred abundances. One approach to limit this effect is to fit the emission measure and the abundance of each element locally, i.e. within a narrow energy band centred on its K-shell transitions, while keeping all the other parameters fixed to their best broad-band fit values \citep[e.g.][]{mernier2015,mernier2016a}.

(ii) Background-related issues may also significantly affect the measurements. This is especially true in the outskirts, where a slightly incorrect background subtraction may dramatically bias the temperature, hence the abundances \citep[e.g.][]{deplaa2007}. On the contrary of point sources, the X-ray emission of most low-redshift clusters covers the entire detector field-of-view; it is thus impossible to extract a pure background spectrum within the immediate neighbourhood of a diffuse source like the ICM. Because every selected ICM observed region is basically an addition of source and background counts, the best way to deal with background-related issues is probably to model all its individual components in the spectra \citep[e.g.][]{bulbul2012a,mernier2015,ezer2017}. Details on how to best implement this approach differs between instruments and/or missions, as all of them are not equally sensitive to the same components \citep[for the case of \textit{XMM-Newton} EPIC, see e.g.][]{snowden2008}. For abundance studies (among many other aspects) of the ICM, it is essential for future missions to understand and reduce the instrumental background as much as possible.

(iii) Finally, SN yield models also constitute an additional source of bias if one wants to compare them with ICM abundance ratios (Sect. \ref{sec:SN_models}). Indeed, these models suffer from uncertainties as predicted yields calculated by different groups with very similar assumption may sometimes vary by a factor of 2 \citep[e.g.][]{wiersma2009}. As a consequence, derived estimates such as the relative contribution of SNIa to the total enrichment should be interpreted very carefully \citep{degrandi2009}.


\section{Future missions and concluding remarks}
\label{sec:conclusion}

As we have seen throughout this review, a better understanding of the chemical history of the largest gravitationally bound structures of the Universe would require more accurate abundance measurements. Apart from the systematic uncertainties discussed in Sect.~\ref{sec:biases}, the main limitation we currently encounter is the moderate spectral resolution of CCD instruments on board operational X-ray missions. Grating spectrometers like \textit{XMM-Newton} RGS provide a better resolution of the Fe-L complex, but are limited by other factors such as (i) a relatively narrow spectral window, (ii) the subsequent difficulty to estimate correctly the continuum level (which is essential for deriving absolute abundances), and (iii) the instrumental broadening of the lines due to the spatial extent of the source.

Undoubtedly, the next achievement in accuracy of measuring abundances is done via micro-calorimeters. Although, regrettably, \textit{Hitomi} could only observe the Perseus cluster before its loss of contact, the success of in-flight micro-calorimeters in measuring ICM abundances has been remarkably demonstrated \citep[][Sects.~\ref{sec:metals_history} and \ref{sec:SN_models_Hitomi}]{THC2017}. The re-flight mission \textit{XRISM} \citep{tashiro2018} will pursue these achievements as it will include the same SXS instrument as on board \textit{Hitomi}. In this respect, unveiling the Fe-L complex at a few eV resolution will be invaluable to (i) better understand all the radiative processes at play in the ICM and, consequently, improve even further the spectral codes used to fit the spectra, (ii) alleviate many biases and degeneracies which may affect the Fe abundance in groups and ellipticals, and (iii) dramatically improve the accuracies of specific abundances, such as O, Ne, or Mg.

\begin{figure*}
  \includegraphics[width=1.0\textwidth]{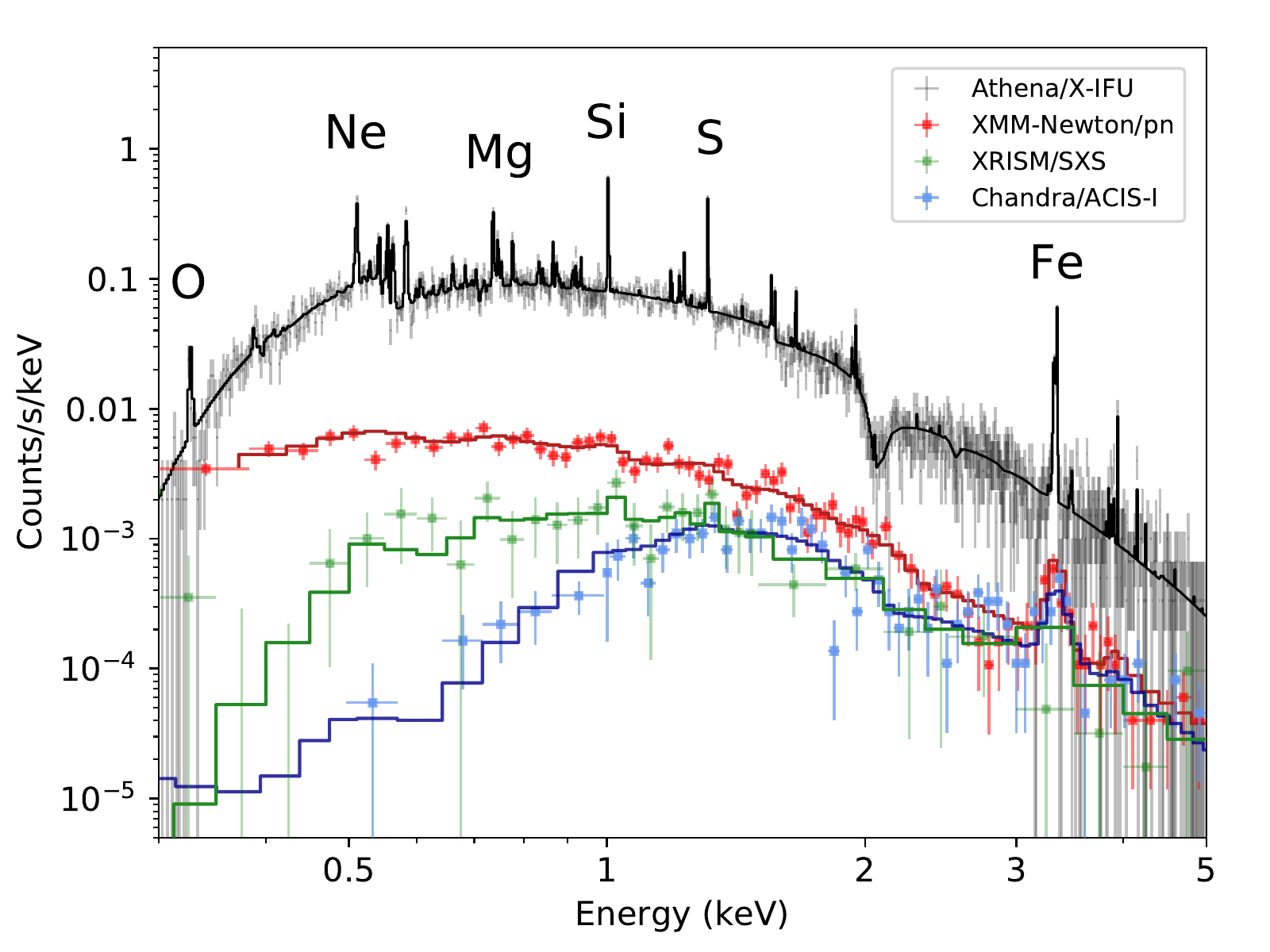}
\caption{Simulated 250 ks spectrum of the core of a distant cluster ($kT = 3$ keV, $z=1$) with the \textit{Athena} X-IFU instrument. For comparison, similar simulated spectra are also shown for the \textit{XMM-Newton} pn, \textit{Chandra} ACIS-I, and \textit{Hitomi}/\textit{XRISM} SXS instruments. For clarity, the data points have been rebinned.}
\label{fig:Athena_sim}       
\end{figure*}

Whereas \textit{XRISM} will constitute a major breakthrough in our knowledge of nearby systems, its spatial resolution ($\sim$1.2 arcmin of point spread function) and the effective area of SXS (comparable to \textit{XMM-Newton} RGS at 1 keV) makes this observatory not optimised for studying higher-$z$ systems. On the contrary, the \textit{Athena} observatory is expected to have an effective area of about one order of magnitude larger than those of the current X-ray missions. Its X-ray Integral Field Unit (X-IFU) instrument will have a spatial resolution comparable to the \textit{XMM-Newton} EPIC instruments, with a spectral resolution of $\sim$2.5 eV \citep{barret2018}. It will allow to study high-$z$, still forming clusters and groups and probably to provide the first direct evidence of the early enrichment scenario and to derive abundance maps in lower redshift clusters with unprecedented accuracy, even in the outskirts \citep{pointecouteau2013,cucchetti2018}. To illustrate the potential capabilities of \textit{Athena}, we compare in Fig.~\ref{fig:Athena_sim} mock spectra of a $z=1$ cluster (with $kT = 3$ keV and proto-solar abundances) simulated for 250 ks of exposure, convolved successively with the ACIS-I (\textit{Chandra}), EPIC pn (\textit{XMM-Newton}), SXS (\textit{XRISM}), and X-IFU (\textit{Athena}) instruments. Even at such distance, it will also be possible to significantly detect and constrain the abundances of elements other than Fe.

In order to enable \textit{XRISM} and \textit{Athena} to push our understanding of the evolution of the ICM enrichment to the next level, it is absolutely necessary that some of the current systematic limitations are reduced and better understood in parallel. This includes, for instance, improvements in:

\begin{enumerate}
\item spectral codes and atomic data;
\item stellar and SNIa/SNcc nucleosynthesis models and yields;
\item hydrodynamical simulations and their convergence in key predictions that will eventually be observable \citep[][in this topical collection]{biffi_review}.
\end{enumerate}

Assuming that the synergy between instrumental and theoretical improvements will be pursued and even strengthened, the future of abundances measurements in the ICM (and of their astrophysical interpretation) will be bright.

\begin{acknowledgements}
We thank the anonymous referee for his/her valuable comments which helped to improve this review. This work was supported by the Lend\"ulet LP2016-11 grant awarded by the Hungarian Academy of Sciences.
P.M. acknowledges support from Russian Science Foundation (grant 14-22-00271).
A.S. is grateful for the support from the Women In Science Excel (WISE) programme of the NWO, and thanks the Kavli Institute for the Physics and Mathematics of the Universe for their continued hospitality. 
S.E. acknowledges financial contribution from the contracts NARO15 ASI-INAF I/037/12/0, ASI 2015-046-R.0 and ASI-INAF n.2017-14-H.0.
SRON is supported financially by NWO, the Netherlands Organization for Scientific Research.

\end{acknowledgements}

\bibliographystyle{aps-nameyear}      
\bibliography{review_abundances}                
\nocite{*}

\end{document}